\documentclass[12pt,french,english]{article}
\usepackage[T1]{fontenc}
\usepackage[latin9]{inputenc}
\usepackage{textcomp}
\usepackage{amsmath}
\usepackage{color}
\usepackage{graphicx}
\usepackage{amssymb}

\newcommand{\lyxline}[1][1pt]{%
  \par\noindent%
  \rule[.5ex]{\linewidth}{#1}\par}

\usepackage{a4wide}
\usepackage{color}
\def\defi{\stackrel{\rm def}{=}}
\usepackage{graphics}
\usepackage{floatflt}
\usepackage{amsmath, amsthm, amssymb}

\usepackage{babel}
\addto\extrasfrench{\providecommand{\fg}{\ifdim\lastskip>\z@\unskip\fi~\frqq}}

\begin{document}
\newcommand{\np}{\newpage}

\newcommand{\vs}{\vspace{1cm}}

\newcommand{\K}{\mathbb {K}}

\newcommand{\bs}{\blacksquare}

\newtheorem{prop}{Proposition} 

\newtheorem{theo}{Theorem} 
\newtheorem{coro}{Corollary} 

\newtheorem{lem}{Lemma} 
\newtheorem{example}{Example} 
\newtheorem{conj}{Conjecture}

\newtheorem{Def}{Definition}

\newcommand{\bp}{\begin{prop}}

\newcommand{\ep}{\end{prop}}

\title{Semi-classical approach for Anosov diffeomorphisms and Ruelle resonances}

\author{Frédéric Faure\textit{}%
\thanks{Institut Fourier, 100 rue des Maths, BP74 38402 St Martin d'Hères.
frederic.faure@ujf-grenoble.fr http://www-fourier.ujf-grenoble.fr/\textasciitilde{}faure%
}, Nicolas Roy%
\thanks{Geometric Analysis group, Institut fr Mathematik, Humbold Université,
Berlin. roy@math.hu-berlin.de%
}, Johannes Sjöstrand%
\thanks{CMLS, Ecole Polytechnique, FR 91128 Palaiseau cedex. johannes@math.polytechnique.fr%
}}
\maketitle
\begin{abstract}
In this paper, we show that some spectral properties of Anosov diffeomorphisms
can be obtained by semi-classical analysis. In particular the Ruelle
resonances which are eigenvalues of the Ruelle transfer operator acting
in suitable anisotropic Sobolev spaces and which govern the decay
of dynamical correlations, can be treated as the quantum resonances
of open quantum systems in the Aguilar-Baslev-Combes theory or the
more recent Helffer-Sjöstrand phase-space theory \cite{sjostrand_87}.

\end{abstract}
\footnote{preprint: http://arxiv.org/abs/0802.1780

transfer operators, and other functional analytic techniques in dynamical
systems 81Q20 Semi-classical techniques

}

\lyxline{\normalsize}

\tableofcontents{}

\lyxline{\normalsize}

\section{Introduction}

An Anosov diffeomorphism $f$ on a compact manifold $M$ is characterized
by the fact that under iterations, every trajectory has hyperbolic
instability, which means that two points that are close to each other
will be separated exponentially fast under the dynamics in the future
or in the past. As a consequence the behavior of individual trajectories
looks like unpredictable or {}``chaotic''. Instead of looking at
individual trajectories, it is then more natural to study a set of
trajectories, or equivalently the transport of functions (or densities)
under the map. One is led to study the so-called Ruelle transfer operator
$\hat{F}$ defined by $\hat{F}\varphi=\varphi\circ f$ with $\varphi\in C^{\infty}\left(M\right)$.
The spectral decomposition of this operator provides objects invariant
under the dynamics and therefore informs us on the long time behavior
of the dynamics, such as ergodicity, mixing, decay of correlations,
central limit theorem ...(see \cite[chap. VII]{reed-simon1},\cite{baladi_livre_00,gaspard-98,katok_hasselblatt}).

The main subject of this paper is the spectral properties of the transfer
operator $\hat{F}$. The approach we propose is based on an elementary
but crucial observation which itself relies on the hypothesis of hyperbolicity:
high Fourier modes of $\hat{F}^{n}\varphi$ go towards infinity as
$n\rightarrow\infty$ or $n\rightarrow-\infty$. In other words, the
variations of the function $\varphi$ evolve towards finer and finer
scales as $n\rightarrow\pm\infty$, and as a consequence the {}``information''
about the initial function $\varphi$ disappear from the macroscopic
scale (the observation scale). This is the mechanism responsible for
chaotic behavior, and in particular for the decay of dynamical correlation
functions. In the 70's, David Ruelle has initiated a fruitful theory
called thermodynamic formalism \cite{ruelle_75,eckmann_ruelle_85,ruelle_86}
where he studied the transfer operator $\hat{F}$ and defined the
Ruelle resonances which govern the exponential decay of the dynamical
correlation functions. This approach has recently been improved considerably
in the works of M. Blank, S. Gouëzel, G. Keller, C. Liverani \cite{liverani_02,liverani_04,liverani_05}
and V. Baladi and M. Tsujii \cite{baladi_sobolev_05,baladi_05} (see
\cite{baladi_05} for some historical remarks) where the authors demonstrate
that the Ruelle resonances are the discrete spectrum of the transfer
operator in suitably defined functional spaces.

From a mathematical point of view, the escape of the function $\hat{F}^{n}\varphi$
towards high Fourier modes we are interested in, is similar to the
escape of a quantum wave function towards infinity in space occurring
in open quantum systems. In such systems, studied since a long time
because of their relevance to spontaneous emission of light in atoms
\cite{cohen3} or radioactive decay in nuclei, physicists and mathematicians
have elaborated concepts and techniques. In the 70's, J. Aguilar,
E. Balslev, J.M. Combes, B. Simon and others developed a mathematical
theory for quantum resonances, which has been improved after by many
authors in the 80's, in particular B. Helffer and J. Sjöstrand \cite{sjostrand_87,sjostrand_resonances_02}.

The aim of this paper is to show that some results of C. Liverani
et al. \cite{liverani_02,liverani_04,liverani_05,liverani_tsujii_06},
V. Baladi et al.~\cite{baladi_sobolev_05,baladi_05} concerning the
definition and properties of Ruelle resonances fit perfectly well
within the semi-classical approach of quantum resonances in phase
space developed in \cite{sjostrand_87,sjostrand_resonances_02,sjoestrand_07}.
The results we present are not new, but we want to show the relevance
of the semi-classical analysis to the theory of hyperbolic dynamical
systems.

Semi-classical analysis (or equivalently microlocal analysis) has
been developed for the study of partial differential equations in
the regime of small wave-length or equivalently, high Fourier mode
regime \cite{martinez-01,dimassi-99,zworski-03}. As we explained
above, the very definition of hyperbolic dynamics involves high Fourier
modes and this implies that semi-classical analysis should be a {}``natural''
approach for its understanding. The idea of relevance of semi-classical
analysis in the context of hyperbolic dynamics has been presented
and used in \cite{fred-RP-06} in a simpler framework (i.e. real analytical
maps on the torus). In this paper we present the later approach in
wider generality.

In semi-classical analysis we distinguish two kinds of operators,
the pseudo-differential operators (PDO) and the Fourier integral operators
(FIO). To each PDO $\hat{P}$ is associated a function $P=\sigma\left(\hat{P}\right)$
on the cotangent bundle $T^{*}M$, called its symbol. To each FIO
$\hat{F}$ is associated a symplectic map $F$ on $T^{*}M$. In semi-classical
analysis, we manipulate the symbols instead of the operators, and
powerful theorems transcribe properties of the symbols in terms of
properties of the operators (for example spectral properties). 

In our context, the Ruelle transfer operator $\hat{F}$ is a FIO whose
associated symplectic map denoted $F:T^{*}M\rightarrow T^{*}M$, is
the lift of $f^{-1}$ (the inverse of the Anosov diffeomorphism).
This is presented in Section 2.

In Section 3, we study the dynamics of $F$. It appears that in $T^{*}M$,
the trajectories of $F$ are non compact, except for the maximal compact
invariant subspace, the section $\xi=0$, and this is related in an
essential way to the discreteness of the spectrum of the operator
$\hat{F}$ obtained after. We construct an {}``escape function''
$A_{m}$ on the cotangent space $T^{*}M$, which decreases strictly
along the non-compact trajectories of $F$, in a controlled manner.
Since it decreases in the unstable direction and increases in the
stable direction, the escape function $A_{m}$ belongs to a class
of symbols with variable order, defined in Appendix A. We defined
an associated invertible PDO denoted by $\hat{A}_{m}$. We also define
the anisotropic Sobolev space associated to $\hat{A}_{m}$ in the
standard manner: $H^{m}\defi\hat{A}_{m}^{-1}\left(L^{2}\left(M\right)\right)$.

In Section 4, we show in Theorem \ref{thm:Ruelle-resonances} that
the operator $\hat{F}$ acting on the anisotropic Sobolev space $H^{m}$
has a discrete spectrum outside an a disk of radius $\varepsilon_{m}$
(which can be made arbitrary small). The discrete spectrum does not
depend on the choice of $A_{m}$ and defines the Ruelle resonances.
This is the main result of this paper. This theorem has already been
obtained by various authors with different degrees of generalities
\cite{liverani_04,liverani_05,baladi_sobolev_05,liverani_tsujii_06},
but the proof we present here is different as it uses in a simple
way three major Theorems of semi-classical analysis: the {}``Composition
Theorem for PDO'', the {}``Egorov's Theorem for transport'' and
the {}``$L^{2}$ continuity Theorem''. 

As an application of this approach, we derive expressions for dynamical
correlation functions in Section 5. In Section 6, we propose a new
proof for a theorem of D. Anosov which states that an Anosov diffeomorphism
preserving a smooth measure is mixing. In Section 7, we show that
a semi-classical truncation of the operator $\hat{F}$ gives the Ruelle
resonance spectrum. This latter result is useful for numerical computations. 

In appendix A, we provide a self-contained presentation of semi-classical
results adapted for this article.

The case of uniformly expanding maps can be considered with a similar
approach. However it would be mostly simplified by the fact that the
escape function $A_{m}$ would have constant order $m$, and the associated
Sobolev space $H^{m}$ are usual (non anisotropic) Sobolev spaces.

\paragraph{Acknowledgment:}

We gratefully acknowledge Mady Smets and {}``Le foyer d'humanisme
de Peyresq'' for their nice hospitality during a workshop where major
part of this work has been made. FF acknowledges support by {}``Agence
Nationale de la Recherche'' under the grant JC05\_52556. We thank
the {}``Classical and quantum resonances team'' Nalini Anantharaman,
Viviane Baladi, Yves Colin de Verdière, Colin Guillarmou, Luc Hillairet,
Frédéric Naud, Stéphane Nonnenmacher and Dominique Spehner for discussions
related to this work.

\section{The model of hyperbolic map}

Let $M$ be a smooth compact connected manifold. Let $f:M\rightarrow M$
be a $C^{\infty}$ Anosov diffeomorphism. We recall the definition:

\vspace{0.cm}\begin{center}{\color{red}\fbox{\color{black}\parbox{16cm}{
\begin{Def}

(see \cite{katok_hasselblatt} page 263) A diffeomorphism $f:M\rightarrow M$
is \textbf{Anosov} (or uniformly hyperbolic) if there exists a Riemannian
metric $g_{0}$, an $f$-\emph{invariant orthogonal} decomposition
of $TM$:\begin{equation}
TM=E_{u}\oplus E_{s}\label{eq:foliation}\end{equation}
and $0<\theta<1$, such that for any $x\in M$\begin{eqnarray}
\left|D_{x}f\left(v_{s}\right)\right|_{g_{0}} & \leq & \theta\left|v_{s}\right|_{g_{0}},\qquad\forall v_{s}\in E_{s}\left(x\right)\label{eq:def_dynamics}\\
\left|D_{x}f^{-1}\left(v_{u}\right)\right|_{g_{0}} & \leq & \theta\left|v_{u}\right|_{g_{0}},\qquad\forall v_{u}\in E_{u}\left(x\right).\nonumber \end{eqnarray}
This means that $E_{s}$ is the stable foliation and $E_{u}$ the
unstable foliation for positive time.

\end{Def}
}}}\end{center}\vspace{0.cm}

\paragraph{Remarks:}
\begin{enumerate}
\item Standard examples are hyperbolic automorphisms of the torus $\mathbb{T}^{n}$
as well as their $C^{1}$ small perturbations, thanks to the structural
stability theorem (see \cite{katok_hasselblatt} page 266). The problem
of classifying manifolds that admit Anosov diffeomorphisms turned
out to be very difficult. The only known examples are infranil manifolds
(which contain the torus case) and it is conjectured that they are
the only ones \cite[p. 16]{hasselblatt_02}. Here is a simple example
of Anosov diffeomorphism on $\left(x,y\right)\in\mathbb{T}^{2}=\mathbb{R}^{2}/\mathbb{Z}^{2}$:\begin{equation}
f:\left(\begin{array}{c}
x\\
y\end{array}\right)\longmapsto\left(\begin{array}{c}
x'\\
y'\end{array}\right)=\left(\begin{array}{cc}
2 & 1\\
1 & 1\end{array}\right)\left(\begin{array}{c}
x\\
y\end{array}\right)+\left(\begin{array}{c}
0\\
\frac{\varepsilon}{2\pi}\sin\left(2\pi\left(2x+y\right)\right)\end{array}\right)\label{eq:example_on_T2}\end{equation}
with $\varepsilon$ small enough%
\footnote{This example preserves area $dx\wedge dy$.%
}. The Ruelle resonances of this map are depicted on figure \ref{fig:Ruelle-Resonances-of}
page \pageref{fig:Ruelle-Resonances-of}.
\item The metric $g_{0}\left(x\right)$ (called the Lyapounov metric) and
the distributions $E_{u}\left(x\right),E_{s}\left(x\right)\subset T_{x}M$
are in general only Hölder continuous with respect to $x\in M$ (See
\cite{katok_hasselblatt} chap. 19). For the purpose of semi-classical
analysis, one needs a \emph{smooth} metric in order to construct suitable
symbols. In this paper, we will assume that $M$ is endowed with a
smooth Riemannian metric $g$ satisfying \begin{equation}
\frac{1}{c}g_{0}\leq g\leq c.g_{0}\label{eq:metric_compatible}\end{equation}
 uniformly on $M$, with \begin{equation}
1\leq c<\theta^{-\frac{1}{2}}.\label{eq_condition_c}\end{equation}
With $\theta_{*}\defi c^{2}\theta$, this implies that $0<\theta_{*}<1$
and for any $v_{s}\in E_{s}\left(x\right)$ one has estimates similar
to Eq.(\ref{eq:def_dynamics}), but with the metric $g$:\begin{equation}
\left|D_{x}f\left(v_{s}\right)\right|_{g}\leq c\left|D_{x}f\left(v_{s}\right)\right|_{g_{0}}\leq c\theta\left|v_{s}\right|_{g_{0}}\leq c^{2}\theta\left|v_{s}\right|_{g}=\theta_{*}\left|v_{s}\right|_{g}\label{eq:estimate_g}\end{equation}
Similarly for $v_{u}\in E_{u}\left(x\right)$, $\left|D_{x}f^{-1}\left(v_{u}\right)\right|_{g}\leq\theta_{*}\left|v_{u}\right|_{g}$.
Unless specified, we will always work with this metric $g$, which
can be obtained from $g_{0}$ by smoothing%
\footnote{If however the metric $g$ is already given, one can always fulfill
Eq.(\ref{eq_condition_c}) by taking some positive power $f^{n_{0}}$,
$n_{0}\in\mathbb{N}$, of the Anosov map.%
}.
\end{enumerate}

\subsection{Transfer operators}

We denote by $dx=d\mu_{Leb}$ an arbitrary%
\footnote{We have chosen a density $dx$ in order to define $L^{2}\left(M\right)$.
However this choice does not play any role for the principal results
of this paper.%
} smooth density normalized by $\mu_{Leb}\left(M\right)=1$. Let us
define the bounded operator $\hat{F}$ on $L^{2}\left(M\right)$ by%
\footnote{It would have been more natural to consider $\hat{F}\varphi\defi\varphi\circ f^{-1}$
instead, but our choice here follows the paper \cite{fred-RP-06}
where we considered expanding maps which are not invertible. %
}

\begin{equation}
\boxed{\hat{F}\varphi\defi\varphi\circ f,\qquad\varphi\in L^{2}\left(M\right)}\label{eq:def_F^}\end{equation}
called the \textbf{Ruelle transfer operator} (or Koopman operator).

Let us emphasize that in general $f$ does not preserve any smooth
measure, but if $f$ preserves the Lebesgue measure $\mu_{Leb}$ then
$\hat{F}$ is unitary in $L^{2}\left(M\right)$.

Let us remark that the $L^{2}-$adjoint operator $\hat{F}^{*}$ is
given by \[
\left(\hat{F}^{*}\varphi\right)\left(y\right)=\left(\varphi\circ f^{-1}\right)\left(y\right)\left|D_{f^{-1}\left(y\right)}f\right|^{-1}\]
 with $\psi,\varphi\in C^{\infty}\left(M\right)$. The adjoint operator
$\hat{F}^{*}$ also called the \textbf{Perron-Frobenius operator}
is usually considered since it transports densities \cite{liverani_05}.
Our main result, Corollary \ref{cor:quasi-compact} page \pageref{cor:quasi-compact},
concerns the spectrum of both $\hat{F}$ and $\hat{F}^{*}$.

\paragraph{Other types of transfer operators.}

In the context of thermodynamic formalism of dynamical systems introduced
by D. Ruelle et al. (see \cite{zinsmeister_00} chap. 4, \cite{keller_98}
chap. 6), a more general class of transfer operators than Eq.(\ref{eq:def_F^})
is considered and defined as follow. Let $V\in C^{\infty}\left(M\right)$
be a smooth (real or complex) valued function called the potential
and let $\hat{F}_{V}:C^{\infty}\left(M\right)\rightarrow C^{\infty}\left(M\right)$
be defined by\begin{equation}
\left(\hat{F}_{V}\varphi\right)\left(x\right):=\varphi\left(f\left(x\right)\right)e^{V\left(x\right)}\label{eq:def_F_V}\end{equation}

Compared to the simplest case $V=0$, given in Eq.(\ref{eq:def_F^}),
the new term $e^{V}$ is a pseudodifferential operator of order $0$
(as defined in the subsequent sections and in Appendix \ref{Appendix:PDO's-with-variable}).
Consequently the canonical map $F:T^{*}M\rightarrow T^{*}M$ associated
to $\hat{F}_{V}$, Eq.(\ref{eq:Def_F}), is unchanged. This implies
that our main results, Theorem \ref{thm:Quasi-compact} page \pageref{thm:Quasi-compact},
Corollary \ref{cor:quasi-compact} page \pageref{cor:quasi-compact}
and their proof, are the same%
\footnote{Of course the spectrum of $\hat{F}_{V}$ depends on $V$ and the value
$1$ is in general no more an eigenvalue of $\hat{F}_{V}$. Corollary
\ref{cor:spectral_radius_1} and \ref{cor:mixing} page \pageref{cor:spectral_radius_1}
are specific to $\hat{F}_{V=0}$.%
} for this new operator $\hat{F}_{V}$.

There is even a slightly more general class of transfer operators
acting on sections of line bundles%
\footnote{This works also for vector bundles.%
}, for which our results work as well. Since these transfer operators
have interesting connections with quantum chaos and geometric quantization
(see \cite{fred-PreQ-06}) we mention them. A general definition proceeds
as follow.

\selectlanguage{french}%
\vspace{0.cm}\begin{center}{\color{red}\fbox{\color{black}\parbox{16cm}{
\begin{Def}

\selectlanguage{english}%
Let $L\rightarrow M$ be a smooth complex line bundle over $M$. A
\textbf{transfer operator} $\hat{F}$ associated to the smooth diffeomorphism
$f:M\rightarrow M$ is a linear map acting on smooth sections, $\hat{F}:C^{\infty}\left(M;L\right)\rightarrow C^{\infty}\left(M;L\right)$,
such that for any smooth function $\psi\in C^{\infty}\left(M\right)$
and any smooth section $s\in C^{\infty}\left(M;L\right)$ one has\begin{equation}
\left(\hat{F}\left(\psi s\right)\right)=\left(\psi\circ f\right).\left(\hat{F}s\right)\label{eq:def_F_bundle}\end{equation}
One also requires that for any $x\in M$ and $s\in C^{\infty}\left(M;L\right)$,
\begin{equation}
\left(s\circ f\right)\left(x\right)\neq0\Rightarrow\left(\hat{F}s\right)\left(x\right)\neq0\label{eq:non_vanish}\end{equation}

\selectlanguage{french}%
\end{Def}
}}}\end{center}\vspace{0.cm}

\selectlanguage{english}%
This definition of transfer operator generalizes Eq.(\ref{eq:def_F_V}),
since in the case where $L$ is a trivial line bundle, sections are
identified with complex functions thanks to a global non vanishing
section $r\in C^{\infty}\left(M;L\right)$: any global section $s\in C^{\infty}\left(M;L\right)$
can be written $s=\varphi r$ with $\varphi\in C^{\infty}\left(M\right)$.
Let $e^{V}\in C^{\infty}\left(M\right)$ be defined by $\hat{F}\left(r\right)=e^{V}r$
which is possible from (\ref{eq:non_vanish}). Then (\ref{eq:def_F_bundle})
gives $\hat{F}\left(\varphi r\right)=\left(\varphi\circ f\right)e^{V}r$
which is equivalent to (\ref{eq:def_F_V}).

In this paper we will only consider the simpler expression Eq.(\ref{eq:def_F^}).

\subsection{Dynamics on the cotangent space}

\begin{figure}
\begin{centering}
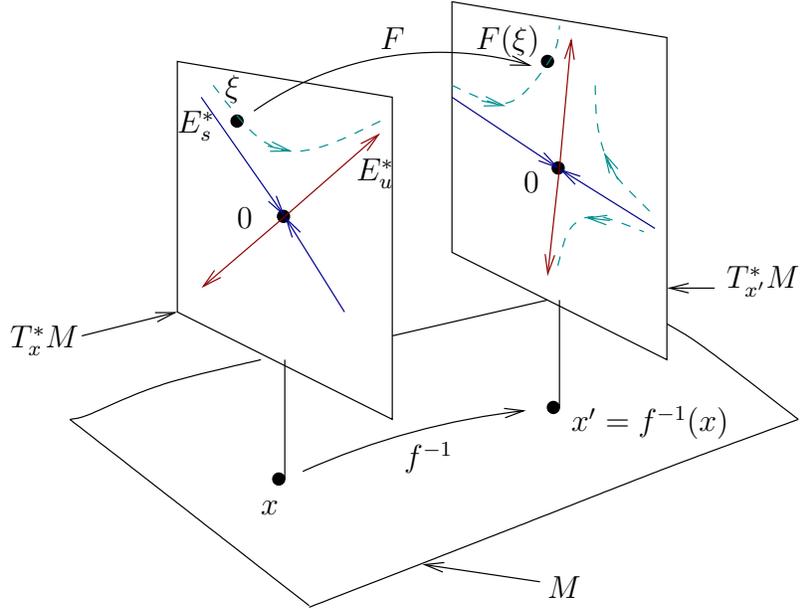
\par\end{centering}

\caption{\label{fig:Dynamics-of-F}Dynamics of $F$ defined in Eq.(\ref{eq:Def_F}),
on the cotangent space $T^{*}M$.}

\end{figure}

In order to study the spectrum of the operator $\hat{F}$ using semi-classical
analysis later on, we need to consider the dynamics induced by $f$
in the cotangent bundle $F:T^{*}M\rightarrow T^{*}M$, namely the
lift of $f^{-1}$. See Figure \ref{fig:Dynamics-of-F}. For any $x\in M$,
let $x'=f^{-1}\left(x\right)$, and define \begin{equation}
\begin{array}{ccc}
F:\quad T_{x}^{*}M & \rightarrow & T_{x'}^{*}M\\
\xi & \longmapsto & \left(D_{x'}f\right)^{t}\xi\end{array}\label{eq:Def_F}\end{equation}

In semi-classical analysis, the map $F$ is precisely the canonical
map associated to the operator $\hat{F}$ defined in Eq.(\ref{eq:def_F^}).
This appears in Egorov's Theorem \ref{thm_egorov}. It is the lift
of $f^{-1}$ rather than $f$ because the support of $\hat{F}\varphi=\varphi\circ f$
is the support of $\varphi$ transported by $f^{-1}$.

Notice that the zero section $\xi=0$ is a compact set, invariant
by the dynamics and its complement contains only unbounded trajectories.
This observation is at the origin of the method which leads to the
quasi-compacity result obtained in the main Theorem \ref{thm:Quasi-compact}.

Let $T^{*}M=E_{s}^{*}\oplus E_{u}^{*}$ be the decomposition dual
to Eq.(\ref{eq:foliation}), i.e. $E_{s}^{*}\left(E_{u}\right)=0\mbox{ and }E_{u}^{*}\left(E_{s}\right)=0$. 

\vspace{0.cm}\begin{center}{\color{blue}\fbox{\color{black}\parbox{16cm}{
\begin{lem}

\label{lem_adjoint_map_anosov}The decomposition $T^{*}M=E_{s}^{*}\oplus E_{u}^{*}$
is invariant by $F$ and: \begin{equation}
\begin{array}{cc}
\left|F\left(\xi_{s}\right)\right|\leq\theta_{*}\left|\xi_{s}\right| & \forall\xi_{s}\in E_{s}^{*}\\
\left|F^{-1}\left(\xi_{u}\right)\right|\leq\theta_{*}\left|\xi_{u}\right| & \forall\xi_{u}\in E_{u}^{*}.\end{array}\label{eq:Inequality_F}\end{equation}
with $\theta_{*}=c^{2}\theta,$ $0<\theta^{*}<1$, with $c$ from
Eq. (\ref{eq:metric_compatible}).

\end{lem}
}}}\end{center}\vspace{0.cm}

\begin{proof}
\small 

The distribution $E_{s}^{*}$ is invariant by $F$ because for any
$\xi_{s}\in E_{s}^{*}$, $v_{u}\in E_{u}$, one has $F\left(\xi_{s}\right)\left(v_{u}\right)=\left(\left(Df\right)^{t}\xi_{s}\right)\left(v_{u}\right)=\xi_{s}\left(Df\left(v_{u}\right)\right)=0$
since $E_{u}$ is invariant by $Df$. The same holds for $E_{u}^{*}$.
On the other hand, one gets easily convinced that Eq. (\ref{eq:metric_compatible})
implies the same inequalities for the metric on the dual space. Eq.(\ref{eq:def_dynamics})
implies that on the dual space, for any $\xi_{s}\in E_{s}^{*}\left(x\right)$,
$\left|F\left(\xi_{s}\right)\right|_{g_{0}}\leq\theta\left|\xi_{s}\right|_{g_{0}}$.
Then for any $\xi_{s}\in E_{s}^{*}\left(x\right)$ one has

\[
\left|F\left(\xi_{s}\right)\right|_{g}\leq c\left|F\left(\xi_{s}\right)\right|_{g_{0}}\leq c\theta\left|\xi_{s}\right|_{g_{0}}\leq c^{2}\theta\left|\xi_{s}\right|_{g}\]
and similarly for $\left|F^{-1}\left(\xi_{u}\right)\right|_{g}\leq c^{2}\theta\left|\xi_{u}\right|,\forall\xi_{u}\in E_{u}^{*}$.

\end{proof}
\normalsize

\section{The escape function and the anisotropic Sobolev spaces}

\subsection{Construction of the escape function $A_{m}$ and the pseudodifferential
operator $\hat{A}_{m}$\label{sub:The-escape-function}}

In this section, we construct a function $A_{m}$ on the cotangent
space which decreases along all the unbounded trajectories of $F$
pictured in Figure \ref{fig:Dynamics-of-F}. It is called an \textbf{escape
function}. In order to apply semi-classical theorems later on, we
make sure that $A_{m}$ is a suitable symbol. This will allow us to
construct a pseudodifferential operator $\hat{A}_{m}$ from the symbol
$A_{m}$. It turns out that an escape function $A_{m}\left(x,\xi\right)$
suitable for our purposes must have an order $m$ in $\left\langle \xi\right\rangle $
which depends itself of the direction $\xi/\left|\xi\right|$. This
gives rise to general classes of symbols $S_{\rho}^{m\left(x,\xi\right)}$
and PDO's $\Psi_{\rho}^{m\left(x,\xi\right)}$ with variable order
$m\left(x,\xi\right)$. Their definitions and main properties are
summarized in Appendix \ref{Appendix:PDO's-with-variable}.

\begin{figure}
\begin{centering}
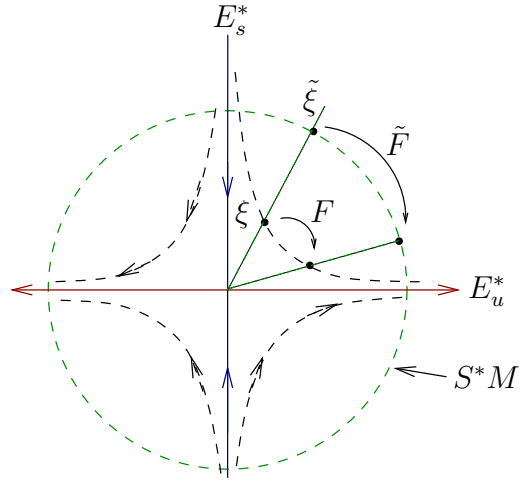
\par\end{centering}

\caption{\label{fig:Cosphere}The map $F$ on the cotangent space $T^{*}M$
and the induced map $\tilde{F}$ on the cosphere bundle $S^{*}M$.}

\end{figure}

\vspace{0.cm}\begin{center}{\color{blue}\fbox{\color{black}\parbox{16cm}{
\begin{lem}

\label{lem:def_A_s_u}Let $u<0<s$. There exists a order function
$m\left(x,\xi\right)\in S_{1}^{0}$ taking values in $\left[u,s\right]$,
with the following properties. For any fixed $x$ and $\left|\xi\right|>1$,
$m\left(x,\xi\right)$ depends only on the direction $\tilde{\xi}=\xi/\left|\xi\right|$
of the cotangent vector. Moreover $m\left(x,\xi\right)=s$ (resp.
$u$) in a vicinity of the stable direction $E_{s}^{*}\left(x\right)$
(resp. unstable direction $E_{u}^{*}\left(x\right)$). See figure
\ref{fig:Cosphere}. The function $m\left(x,\xi\right)$ decreases
with respect to the map $F$: \begin{equation}
\exists R>0,\quad\forall\left|\xi\right|\geq R,\qquad\left(m\circ F\right)\left(x,\xi\right)-m\left(x,\xi\right)\leq0,\label{eq:decrease_of_m}\end{equation}
Define\begin{equation}
A_{m}\left(x,\xi\right)\defi\left\langle \xi\right\rangle ^{m\left(x,\xi\right)},\qquad\left\langle \xi\right\rangle =\sqrt{1+\left|\xi\right|^{2}}.\label{eq:Def_symbol_A}\end{equation}
which belongs to the class $S_{\rho}^{m\left(x,\xi\right)}$, for
any $\frac{1}{2}\leq\rho<1$. The main property of the symbol $A_{m}$
is:\begin{equation}
\exists R>0,\quad\forall\left|\xi\right|\geq R,\qquad\frac{\left(A_{m}\circ F\right)\left(x,\xi\right)}{A_{m}\left(x,\xi\right)}\leq e^{-ca}<1\label{eq:escape_estimate}\end{equation}
with $a=\mbox{min}\left(-u,s\right)$ and $c>0$ independent of the
choice of $u,s$.

\end{lem}
}}}\end{center}\vspace{0.cm}

\paragraph{Remarks:}
\begin{itemize}
\item Eq.(\ref{eq:escape_estimate}) means that the function $A_{m}$ decreases
strictly and uniformly along the trajectories of $F$ in the cotangent
space. We call $A_{m}$ an \textbf{escape function}. 
\item The constancy of $m$ in the vicinity of the stable/unstable direction
allows us to have a smooth order function $m$ despite the foliations
$E_{s}^{*}\left(x\right)$,$E_{u}^{*}\left(x\right)$ have only Hölder
regularity.
\item Inspection of the proof shows that $c$ can be chosen arbitrary close
to $\log\left(\frac{1}{\theta_{*}}\right)$. 
\end{itemize}
The real symbol $A_{m}$ can be quantized into a pseudodifferential
operator $\hat{A}_{m}$ of variable order $m\left(x,\xi\right)$,
according to Eq.(\ref{eq_pdo}). Then Corollary \ref{cor_quantif_aa_inversible}
and Example \ref{exa_elliptic} tell us that we can modify the symbol
$A_{m}$ at a subleading order (i.e. $S_{\rho}^{m\left(x,\xi\right)-\left(2\rho-1\right)}$)
such that the operator can be assumed to be formally \emph{self-adjoint}
and \emph{invertible} on $C^{\infty}\left(M\right)$.

\paragraph{Proof of Lemma \ref{lem:def_A_s_u}.}

\subparagraph{The function $m$.}

Since the lifted map $F$ defined in Eq.(\ref{eq:Def_F}) is linear
in $\xi$, it defines a map $\tilde{F}$ on the cosphere bundle $S^{*}M=\left(T^{*}M\backslash\left\{ 0\right\} \right)/\mathbb{R}^{+}$,
namely the space of directions \[
\tilde{\xi}:=\xi/\left|\xi\right|,\]
which is a compact space. See Figure \ref{fig:Cosphere}. The image
of $E_{u}^{*},E_{s}^{*}\subset T^{*}M$ by the projection $T^{*}M\backslash\left\{ 0\right\} \rightarrow S^{*}M$
are denoted respectively $\tilde{E}_{u}^{*},\tilde{E}_{s}^{*}\subset S^{*}M$.
Eq.(\ref{eq:Inequality_F}) implies that $\tilde{E}_{u}^{*}$ is a
uniform attractor for $\tilde{F}$, and $\tilde{E}_{s}^{*}$ is a
uniform repeller, i.e. $\tilde{F}^{n}\left(\tilde{\xi}\right)$ converges
to $\tilde{E}_{u}^{*}$ (respect. $\tilde{E}_{s}^{*}$) when $n\rightarrow+\infty$
(respect. $n\rightarrow-\infty$). 

Let $u<0<s$. Let $m_{0}\in C^{\infty}\left(S^{*}M;\left[u,s\right]\right)$
with $m_{0}=s>0$ in a neighborhood $\tilde{N}_{s}$ of $\tilde{E}_{s}^{*}$
and $m_{0}=u<0$ in a neighborhood $\tilde{N}_{u}$ of $\tilde{E}_{u}^{*}$.
We also assume that \begin{equation}
(\tilde{\xi}\in\tilde{N}_{s}\Rightarrow\tilde{F}^{-1}\left(\tilde{\xi}\right)\in\tilde{N}_{s})\mbox{ and }(\tilde{\xi}\in\tilde{N}_{u}\Rightarrow\tilde{F}\left(\tilde{\xi}\right)\in\tilde{N}_{u})\label{eq:ee0}\end{equation}

\begin{figure}
\begin{centering}
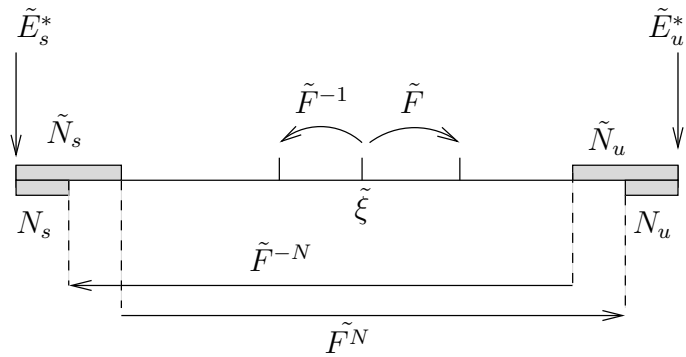
\par\end{centering}

\caption{\label{fig:Cosphere_2}The horizontal axis is a schematic picture
of $S^{*}M$ and this shows the construction and properties of the
sets $N_{s}$ and $N_{u}$.}

\end{figure}

See Figure \ref{fig:Cosphere_2}.%

Let $N\in\mathbb{N}$ and define $\tilde{m}\in C^{\infty}\left(S^{*}M;\left[u,s\right]\right)$
by \begin{equation}
\tilde{m}:=\frac{1}{2N}\sum_{n=-N}^{N-1}m_{0}\circ\tilde{F}^{n}\label{eq:def_m_tilde}\end{equation}
Then\begin{equation}
\tilde{m}\circ\tilde{F}-\tilde{m}=\frac{1}{2N}\left(m_{0}\circ\tilde{F}^{N}-m_{0}\circ\tilde{F}^{-N}\right)\label{eq:m_N}\end{equation}

We will show now that \begin{equation}
\forall\tilde{\xi}\in S^{*}M\qquad\tilde{m}\left(\tilde{F}\left(\tilde{\xi}\right)\right)-\tilde{m}\left(\tilde{\xi}\right)\leq0.\label{eq:e4}\end{equation}

Let $N_{s}:=S^{*}M\setminus\tilde{F}^{-N}\left(\tilde{N}_{u}\right)$
and $N_{u}:=S^{*}M\setminus\tilde{F}^{N}\left(\tilde{N}_{s}\right)$.
Then $m_{0}\left(\tilde{F}^{N}\left(\tilde{\xi}\right)\right)=u$
for $\tilde{\xi}\notin N_{s}$, and similarly $m_{0}\left(\tilde{F}^{-N}\left(\tilde{\xi}\right)\right)=s$
for $\tilde{\xi}\notin N_{u}$.

For $N$ large enough one has $N_{s}\subset\tilde{N}_{s}$, $N_{u}\subset\tilde{N}_{u}$
and $N_{s}\cap N_{u}=\emptyset$. Therefore, 
\begin{itemize}
\item if $\tilde{\xi}\in N_{s}$ then $\tilde{\xi}\notin N_{u}$ and $m_{0}\left(\tilde{F}^{-N}\left(\tilde{\xi}\right)\right)=s\geq m_{0}\left(\tilde{F}^{N}\left(\tilde{\xi}\right)\right)$
and (\ref{eq:m_N}) gives $\tilde{m}\left(\tilde{F}\left(\tilde{\xi}\right)\right)-\tilde{m}\left(\tilde{\xi}\right)\leq0$. 
\item Similarly if $\tilde{\xi}\in N_{u}$ then $\tilde{\xi}\notin N_{s}$
and $m_{0}\left(\tilde{F}^{N}\left(\tilde{\xi}\right)\right)=u\leq m_{0}\left(\tilde{F}^{-N}\left(\tilde{\xi}\right)\right)$
thus $\tilde{m}\left(\tilde{F}\left(\tilde{\xi}\right)\right)-\tilde{m}\left(\tilde{\xi}\right)\leq0$. 
\item Finally if $\tilde{\xi}\notin\left(N_{u}\cup N_{s}\right)$ then $m_{0}\left(\tilde{F}^{N}\left(\tilde{\xi}\right)\right)-m_{0}\left(\tilde{F}^{-N}\left(\tilde{\xi}\right)\right)=\left(u-s\right)<0$
and therefore \begin{equation}
\forall\tilde{\xi}\notin\left(N_{u}\cup N_{s}\right)\qquad\tilde{m}\left(\tilde{F}\left(\tilde{\xi}\right)\right)-\tilde{m}\left(\tilde{\xi}\right)=\frac{1}{2N}\left(u-s\right)<0\label{eq:strict_inf}\end{equation}

\end{itemize}
We have shown Eq.(\ref{eq:e4}).

We construct a smooth function $m$ on $T^{*}M$ satisfying\begin{eqnarray*}
m\left(x,\xi\right) & = & \tilde{m}\left(\tilde{\xi}\right),\qquad\mbox{if}\left|\xi\right|>1,\\
 & = & 0\qquad\mbox{if}\left|\xi\right|<1/2\end{eqnarray*}
Then (\ref{eq:e4}) implies Eq.(\ref{eq:decrease_of_m}).

From Eq.(\ref{eq:ee0}) one deduces that the set\[
\mathcal{S}\left(\tilde{\xi}\right):=\left\{ n\in\mathbb{Z}\quad/\quad\tilde{F}^{n}\left(\tilde{\xi}\right)\notin\left(N_{s}\cup N_{u}\right)\right\} \]
is connected. Moreover the cardinal of this set is uniformly bounded:\[
\exists\mathcal{N}\in\mathbb{N}\quad\forall\tilde{\xi}\in S^{*}M\qquad\sharp\mathcal{S}\left(\tilde{\xi}\right)\leq\mathcal{N}\]

From this we deduce that
\begin{itemize}
\item if $\tilde{\xi}\in N_{s}$ then $\tilde{F}^{N}\left(\tilde{\xi}\right)\notin\tilde{N}_{u}$
but also $\tilde{F}^{n}\left(\tilde{\xi}\right)\notin\tilde{N}_{u}$
for $n\leq N$ and even $\tilde{F}^{n}\left(\tilde{\xi}\right)\in\tilde{N}_{s}$
for $n\leq N-\mathcal{N}$. Thus (\ref{eq:def_m_tilde}) gives\begin{equation}
\tilde{m}\left(\tilde{F}\left(\tilde{\xi}\right)\right)\geq\left(1-\frac{\mathcal{N}+1}{2N}\right)s+\frac{\mathcal{N}+1}{2N}u\geq\frac{s}{2}\label{eq:e2}\end{equation}
where the last inequality holds for $N$ large enough.
\item If $\tilde{\xi}\in N_{u}$ one shows similarly that\begin{equation}
\tilde{m}\left(\tilde{\xi}\right)\leq\left(1-\frac{\mathcal{N}}{2N}\right)u+\frac{\mathcal{N}}{2N}\leq\frac{u}{2}\label{eq:e3}\end{equation}
where the last inequality holds for $N$ large enough.
\end{itemize}

\subparagraph{The symbol $A_{m}$.}

Let \[
A_{m}\left(x,\xi\right):=\left\langle \xi\right\rangle ^{m\left(x,\xi\right)}\]
 with $\left\langle \xi\right\rangle =\sqrt{1+\left|\xi\right|^{2}}$.
$A_{m}$ belongs to the class $S_{\rho}^{m\left(x,\xi\right)}$, for
any $\frac{1}{2}\leq\rho<1$ from Lemma \ref{lem_ksi_puissance_m_is_good_symbol}.
We will show now the uniform escape estimate Eq.(\ref{eq:escape_estimate}).

For $\left|\xi\right|$ large enough one has \[
\log A_{m}\left(x,\xi\right)=\tilde{m}\left(\tilde{\xi}\right)\ln\left\langle \xi\right\rangle ,\qquad\log\left(A_{m}\circ F\right)\left(x,\xi\right)=\tilde{m}\left(\tilde{F}\left(\tilde{\xi}\right)\right)\ln\left\langle F\left(\xi\right)\right\rangle \]

\begin{itemize}
\item If $\tilde{\xi}\notin\left(N_{s}\cup N_{u}\right)$ then (\ref{eq:strict_inf})
gives $\tilde{m}\left(\tilde{F}\left(\tilde{\xi}\right)\right)=\tilde{m}\left(\tilde{\xi}\right)+\frac{1}{2N}\left(u-s\right)$.
One also has $\log\left\langle F\left(\xi\right)\right\rangle =\log\left\langle \xi\right\rangle +\mathcal{O}\left(1\right)$.
Therefore \begin{eqnarray*}
\tilde{m}\left(\tilde{F}\left(\tilde{\xi}\right)\right)\ln\left\langle F\left(\xi\right)\right\rangle -\tilde{m}\left(\tilde{\xi}\right)\ln\left\langle \xi\right\rangle  & = & \left(\tilde{m}\left(\tilde{\xi}\right)+\frac{1}{2N}\left(u-s\right)\right)\left(\log\left\langle \xi\right\rangle +\mathcal{O}\left(1\right)\right)-\tilde{m}\left(\tilde{\xi}\right)\ln\left\langle \xi\right\rangle \\
 & = & \tilde{m}\left(\tilde{\xi}\right)\mathcal{O}\left(1\right)+\frac{1}{2N}\left(u-s\right)\log\left\langle \xi\right\rangle \\
 & \leq & -c\min\left(s,-u\right)\end{eqnarray*}
with $c>0$ and if $\left|\xi\right|$ is large enough.
\item If $\tilde{\xi}\in N_{s}$ (neighborhood of the stable direction)
then $\left|F\left(\xi\right)\right|\leq\frac{1}{C'}\left|\xi\right|$,
with $C'>1$, so $0<\ln\left\langle F\left(\xi\right)\right\rangle \leq\ln\left\langle \xi\right\rangle -\ln C$,
with $C>1$ (close to $C'$). And using (\ref{eq:e2}),(\ref{eq:e4})
\begin{eqnarray*}
\tilde{m}\left(\tilde{F}\left(\tilde{\xi}\right)\right)\ln\left\langle F\left(\xi\right)\right\rangle  & \leq & \tilde{m}\left(\tilde{F}\left(\tilde{\xi}\right)\right)\left(\ln\left\langle \xi\right\rangle -\ln C\right)\\
 & \leq & \tilde{m}\left(\tilde{\xi}\right)\ln\left\langle \xi\right\rangle -\frac{s}{2}\ln C\end{eqnarray*}

\item If $\tilde{\xi}\in N_{u}$ (neighborhood of the unstable direction)
then $\left|F\left(\xi\right)\right|\geq C'\left|\xi\right|$, with
$C'>1$, so $\ln\left\langle F\left(\xi\right)\right\rangle \geq\ln\left\langle \xi\right\rangle +\ln C$
with $C>1$ (close to $C'$). And using (\ref{eq:e2}),(\ref{eq:e4}),\begin{eqnarray*}
\tilde{m}\left(\tilde{F}\left(\tilde{\xi}\right)\right)\ln\left\langle F\left(\xi\right)\right\rangle  & \leq & \tilde{m}\left(\tilde{\xi}\right)\left(\ln\left\langle \xi\right\rangle +\ln C\right)\\
 & \leq & \tilde{m}\left(\tilde{\xi}\right)\ln\left\langle \xi\right\rangle +\frac{1}{2}u\ln C\end{eqnarray*}

\end{itemize}
In conclusion, for any $x$ and $\left|\xi\right|$ large enough,
there exists $c>0$ independent of $u,s$ such that\[
\log\left(A_{m}\circ F\right)\left(x,\xi\right)\leq\log A_{m}\left(x,\xi\right)-c\min\left(s,-u\right)\]

We have obtained the uniform escape estimate Eq.(\ref{eq:escape_estimate})
and finished the proof of Lemma \ref{lem:def_A_s_u}.

\subsection{The anisotropic Sobolev spaces}

A particular feature of the self-adjoint and invertible PDO $\hat{A}_{m}\in\Psi_{\rho}^{m\left(x,\xi\right)}$
introduced above is that its symbol $A_{m}$ has a non-isotropic behavior
with respect to $\xi\in T_{x}^{*}M$. It is a PDO with maximum order
$a=\min\left(\left|u\right|,s\right)$, but with variable order $m\left(x,\xi\right)\in\left[u,s\right]$,
with $u<0<s$. For large $\left|\xi\right|$, the symbol $A_{m}$
increases in the stable direction $\xi\in E_{s}^{*}\left(x\right)$
as $A_{m}\left(\xi\right)\sim\left|\xi\right|^{s}$ and decreases
in the unstable direction $\xi\in E_{u}^{*}\left(x\right)$ as $A_{m}\left(\xi\right)\sim1/\left|\xi\right|^{\left|u\right|}$.
In this section we consider a slightly more general space of functions
$m$ and do not require any relation with the dynamics of $F$: we
just assume that $m$ is in $C^{\infty}\left(T^{*}M\right)$ and is
a function of $\left(x,\frac{\xi}{\left|\xi\right|}\right)$ for $\left|\xi\right|$
large enough. Therefore $m\in S_{1}^{0}$ is an order function. 

We define the \textbf{anisotropic Sobolev space} to be the space of
distributions (included in $\mathcal{D}'\left(M\right)$):\begin{equation}
\boxed{H^{m}\defi\hat{A}_{m}^{-1}\left(L^{2}\left(M\right)\right)}\label{eq:def_Hm}\end{equation}

\paragraph{Remarks:}
\begin{itemize}
\item This definition is very similar to the standard definition of Sobolev
spaces on $\mathbb{R}^{d}$ (\cite{taylor_tome1} p.271): \[
H^{s}\defi Op\left(\left\langle \xi\right\rangle ^{s}\right)^{-1}\left(L^{2}\left(\mathbb{R}^{d}\right)\right)\]
 except for anisotropy with respect to $\xi$. Equivalent definitions
of anisotropic Sobolev spaces have been given in \cite{leopold_91}
and also by V. Baladi et M. Tsujii in \cite{baladi_05} for the specific
purpose of hyperbolic dynamics.
\item Notice that $\varphi\in H^{m}\Leftrightarrow\hat{A}_{m}\varphi\in L^{2}\left(M\right)$,
so roughly speaking, in the case of the function $m$ defined in Lemma
\ref{lem:def_A_s_u}, it means that the Fourier transform $\hat{\varphi}\left(\xi\right)$
performed in a vicinity of $x\in M$, increases less than $\left|\xi_{s}\right|^{-s-d/2}$
for $\xi_{s}\in E_{s}^{*}\left(x\right)$ and less than $\left|\xi_{u}\right|^{-u-d/2}$
for $\xi_{u}\in E_{u}^{*}\left(x\right)$, with $d=\mbox{dim}\left(M\right)$.
We can say that if $u<0<s$, then $\varphi$ is regular in the stable
direction and irregular in the unstable direction.
\end{itemize}

\paragraph{Some simple properties of the anisotropic Sobolev spaces $H^{m}$:}
\begin{enumerate}
\item $H^{m}$ is a Hilbert space with the scalar product \[
\left(\varphi_{1},\varphi_{2}\right)_{H^{m}}\defi\left(\hat{A}_{m}\varphi_{1},\hat{A}_{m}\varphi_{2}\right)_{L^{2}\left(M\right)},\qquad\varphi_{1},\varphi_{2}\in H^{m}\]
 and the map \begin{equation}
\hat{A}_{m}:\left(H^{m},\left(.,.\right)_{H^{m}}\right)\rightarrow\left(L^{2}\left(M\right),\left(.,.\right)_{L^{2}}\right)\label{eq:Unitary_isom}\end{equation}
is unitary.
\item On $L^{2}\left(M\right)$, $\mbox{Dom}\left(\hat{A}_{m}\right)=H^{m}\cap L^{2}\left(M\right)$
and $\mbox{Dom}\left(\hat{A}_{m}^{-1}\right)=H^{-m}\cap L^{2}\left(M\right)$. 
\item There are embedding relations as for usual Sobolev spaces. First\begin{equation}
H^{\max\left(m\right)}\subset H^{m}\subset H^{\min\left(m\right)}\label{eq:Sobolev_embed}\end{equation}
If $m'\geq m$ then\begin{equation}
H^{m'}\subset H^{m}\label{eq:Hm1_in_Hm2}\end{equation}
and $H^{m'}$ is dense in $H^{m}$.
\item If $\varphi\in H^{m}$ and $g\in C^{\infty}\left(M\right)$ then \begin{equation}
g\varphi\in H^{m}\label{eq:f-phi}\end{equation}
and moreover, the map $\varphi\rightarrow g\varphi$ is continuous
$H^{m}\rightarrow H^{m}$.
\item Let \[
H^{-m}\defi\hat{A}_{m}\left(L^{2}\left(M\right)\right).\]
The spaces $H^{m}$ and $H^{-m}$ are dual in the following sense:
if $\psi\in H^{m},\varphi\in H^{-m}$, we note\begin{equation}
\overline{\psi}\left(\varphi\right)=\varphi\left(\overline{\psi}\right)=\left(\psi,\varphi\right)_{H^{m}\times H^{-m}}\defi\left(\hat{A}_{m}\psi,\hat{A}_{m}^{-1}\varphi\right)_{L^{2}\left(M\right)}.\label{eq:def_dualite}\end{equation}
Then\begin{equation}
\left|\left(\psi,\varphi\right)_{H^{m}\times H^{-m}}\right|\leq\left\Vert \psi\right\Vert _{H^{m}}\left\Vert \varphi\right\Vert _{H^{-m}}.\label{eq:Inegalite}\end{equation}

\item If $\psi\in H^{m}\cap L^{2}\left(M\right)$ and $\varphi\in H^{-m}\cap L^{2}\left(M\right)$
then \begin{equation}
\left(\psi,\varphi\right)_{H^{m}\times H^{-m}}=\left(\psi,\varphi\right)_{L^{2}\left(M\right)}\label{eq:Prop_dualite}\end{equation}
Since the dual bracket coincides with the $L^{2}$ scalar product,
we will drop the indices in the sequel of the paper, and write:\[
\left(\psi,\varphi\right)\defi\left(\psi,\varphi\right)_{H^{m},H^{-m}}\]

\item If $\psi\in H^{m}$, $\varphi\in H^{-m}$ and $g\in C^{\infty}$,
then: \begin{equation}
\left(g\psi,\varphi\right)=\left(\psi,\overline{g}\varphi\right)\label{eq:relation}\end{equation}

\end{enumerate}
\begin{proof}
\small 

The proofs of properties 1 to 4 follow directly from those of $\hat{A}_{m}$.

5.$\left|\left(\psi,\varphi\right)\right|=\left|\left(\hat{A}_{m}\psi,\hat{A}_{m}^{-1}\varphi\right)_{L^{2}\left(M\right)}\right|\leq\left\Vert \hat{A}_{m}\psi\right\Vert _{L^{2}}\left\Vert \hat{A}_{m}^{-1}\varphi\right\Vert _{L^{2}}=\left\Vert \psi\right\Vert _{H^{m}}\left\Vert \varphi\right\Vert _{H^{-m}}$.

6.$\left(\psi,\varphi\right)=\left(\hat{A}_{m}\psi,\hat{A}_{m}^{-1}\varphi\right)_{L^{2}\left(M\right)}=\left(\psi,\varphi\right)_{L^{2}\left(M\right)}$
by self adjointness of $\hat{A}_{m}$.

7. Let $\mathcal{M}_{g}$ denotes multiplication $g\in C^{\infty}$.
The operator $\hat{B}_{g}=\hat{A}_{m}\mathcal{M}_{g}\hat{A}_{m}^{-1}$
is bounded in $L^{2}\left(M\right)$ . Moreover, one has $\hat{B}_{g}^{*}=\hat{A}_{m}^{-1}\mathcal{M}_{\overline{g}}\hat{A}_{m}$
since $\hat{A}_{m}$ is self-adjoint. We deduce that \begin{eqnarray*}
\left(g\psi,\varphi\right) & = & \left(\left(\hat{A}_{m}\mathcal{M}_{g}\hat{A}_{m}^{-1}\right)\hat{A}_{m}\psi,\hat{A}_{m}^{-1}\varphi\right)_{L^{2}}=\left(\hat{A}_{m}\psi,\left(\hat{A}_{m}^{-1}\mathcal{M}_{\overline{g}}\hat{A}_{m}\right)\hat{A}_{m}^{-1}\varphi\right)_{L^{2}}\\
 & = & \left(\psi,\overline{g}\varphi\right).\end{eqnarray*}

\end{proof}
\normalsize

\begin{figure}
\begin{centering}
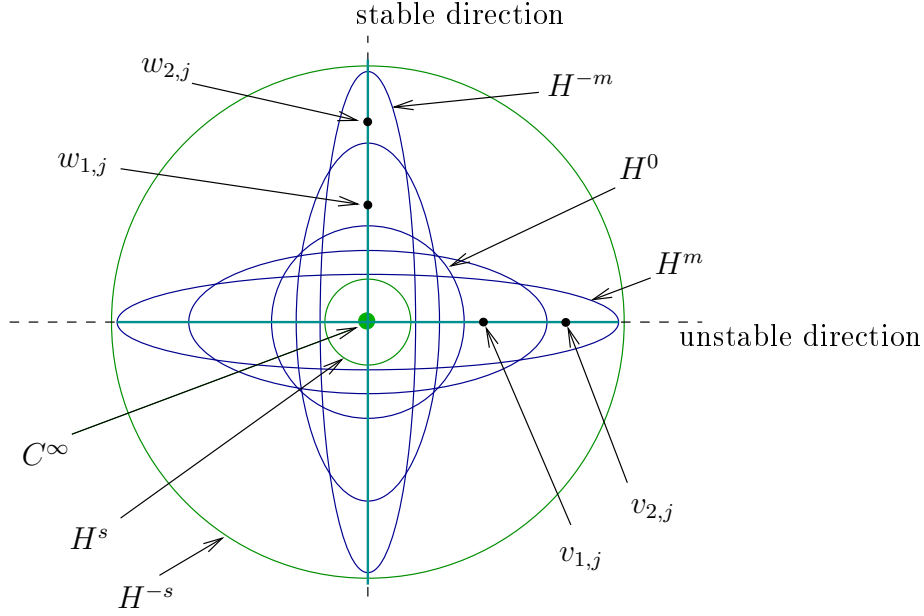
\par\end{centering}

\caption{\label{fig:Sobolev_spaces}Schematic representation of the anisotropic
Sobolev spaces $H^{m}$ and their embedding relations Eq.(\ref{eq:Sobolev_embed})
with the usual Sobolev spaces $H^{s}$, $s\in\mathbb{R}$. The eigen-distributions
$v_{i,j},w_{i,j}$ of the operator $\hat{F}$ which appear in Corollary
\ref{cor:quasi-compact} are also represented.}

\end{figure}

\section{Spectrum of resonances}

We give now the main result of this paper. Its proof relies on semi-classical
analysis and is inspired by the study of resonances in open quantum
systems \cite{sjostrand_87,hislop_95}. It is also inspired from a
previous work \cite{fred-RP-06} performed within a simple and illuminating
model, namely analytical hyperbolic map on the torus. In some sense
it shows a close analogy between Ruelle resonances and quantum resonances.
The essential point of this approach is to find the discrete spectrum
of resonances of the operator $\hat{F}$ in the Sobolev space of distributions
$H^{m}$, thanks to a conjugacy by the escape operator $\hat{A}_{m}$
defined in Section \ref{sub:The-escape-function}. First observe that
the operator $\hat{F}$ defined in Eq.(\ref{eq:def_F^}) extends by
duality to the distribution space $\mathcal{D}'\left(M\right)$ by
\[
\hat{F}\left(\alpha\right)\left(\varphi\right)=\alpha\left(\hat{F}^{*}\left(\varphi\right)\right)\]
where $\alpha\in\mathcal{D}'\left(M\right)$, $\varphi\in C^{\infty}\left(M\right)$
and $\hat{F}^{*}$ is the $L^{2}-$adjoint operator. One checks easily
that for $\psi,\varphi\in L^{2}\left(M\right)$, $\hat{F}^{*}$ is
given by $\left(\hat{F}^{*}\varphi\right)\left(y\right)=\left(\varphi\circ f^{-1}\right)\left(y\right)\left|D_{f^{-1}\left(y\right)}f\right|^{-1}$.

\vspace{0.cm}\begin{center}{\color{blue}\fbox{\color{black}\parbox{16cm}{
\begin{theo}

\label{thm:Quasi-compact}Let $m$ be a function which satisfies the
hypothesis of Lemma \ref{lem:def_A_s_u}. $\hat{F}$ leaves the anisotropic
Sobolev space $H^{m}$ globally invariant. The operator\[
\hat{F}:H^{m}\rightarrow H^{m}\]
is a bounded operator and can be written \begin{equation}
\hat{F}=\hat{r}_{m}+\hat{k}_{m}\label{eq:F_rk}\end{equation}
where $\hat{k}_{m}$ is a compact operator and $\left\Vert \hat{r}_{m}\right\Vert \leq\varepsilon_{m}=\left\Vert \hat{F}\right\Vert _{L^{2}}e^{-ca}$
with constants $c,a>0$ defined in Lemma \ref{lem:def_A_s_u}. Consequently,
the essential spectral radius is smaller than $\varepsilon_{m}$,
which means that $\hat{F}$ has a discrete spectrum $\lambda_{i}$
outside the circle of radius $\varepsilon_{m}$. 

\end{theo}
}}}\end{center}\vspace{0.cm}

\paragraph{Remarks :}
\begin{itemize}
\item $\left\Vert \hat{F}\right\Vert _{L^{2}}$ depends on the choice of
the density $dx$. We could have $\left\Vert \hat{F}\right\Vert _{L^{2}}$
closer to $1$ by another choice of $dx$.
\item Notice that $1$ is an eigenvalue of $\hat{F}$, with constant eigenfunction.
We will see in Corollary \ref{cor:spectral_radius_1} that the spectral
radius of $\hat{F}$ is one, i.e. that there are no eigenvalues outside
the unit circle.
\item For future purpose, let $\varepsilon>0$ and $\mathcal{O}_{\varepsilon}$
denotes the set of order functions $m$ which satisfy the hypothesis
of Lemma \ref{lem:def_A_s_u}, and such that $\varepsilon_{m}=\left\Vert \hat{F}\right\Vert _{L^{2}}e^{-ca}<\varepsilon$:\begin{equation}
\mathcal{O}_{\varepsilon}\defi\left\{ m\quad/\,\varepsilon_{m}=\left\Vert \hat{F}\right\Vert _{L^{2}}e^{-ca}<\varepsilon\right\} \label{eq:def_O_eps}\end{equation}
The set $\mathcal{O}_{\varepsilon}$ is non empty.
\end{itemize}
\begin{proof}
\small 

We use the unitary map between $H^{m}$ and $L^{2}\left(M\right)$
given in Eq.(\ref{eq:def_Hm}), and consider $\hat{Q}_{m}\defi\hat{A}_{m}\hat{F}\hat{A}_{m}^{-1}:L^{2}\left(M\right)\rightarrow L^{2}\left(M\right)$,
defined on a dense domain, and which is unitary equivalent to $\hat{F}:H^{m}\rightarrow H^{m}$:
\[
\begin{array}{ccc}
L^{2}\left(M\right) & \overset{\hat{Q}_{m}}{\rightarrow} & L^{2}\left(M\right)\\
\downarrow\hat{A}_{m}^{-1} & \circlearrowleft & \downarrow\hat{A}_{m}^{-1}\\
H^{m} & \overset{\hat{F}}{\rightarrow} & H^{m}\end{array}\]

Instead of working with $\hat{Q}_{m}$ directly, it is more convenient
to consider \[
\hat{P}_{m}\defi\hat{F}^{-1}\hat{Q}_{m}=\left(\hat{F}^{-1}\hat{A}_{m}\hat{F}\right)\hat{A}_{m}^{-1}.\]
It follows from \emph{Egorov's Theorem} \ref{thm_egorov}, that the
product $\hat{F}^{-1}\hat{A}_{m}\hat{F}$ is a PDO in $\Psi_{\rho}^{m\circ F\left(x,\xi\right)}$
whose symbol is $A_{m}\circ F$ modulo subleading terms in $S_{\rho}^{m\circ F\left(x,\xi\right)-\left(2\rho-1\right)}$.
On the other hand, \emph{the composition Theorem \ref{thm_composition}
for PDO} tells that $\hat{P}_{m}$ is a PDO in $\Psi_{\rho}^{m\circ F\left(x,\xi\right)-m\left(x,\xi\right)}$
whose symbol is $P_{m}=\frac{A_{m}\circ F}{A_{m}}$ modulo subleading
corrections in $S_{\rho}^{m\circ F\left(x,\xi\right)-m\left(x,\xi\right)-\left(2\rho-1\right)}$.
From the construction of the escape function $A_{m}$, Eq.(\ref{eq:decrease_of_m})
insures that $\hat{P}_{m}\in\Psi_{\rho}^{0}$. On the other hand Eq.(\ref{eq:escape_estimate})
gives \[
\limsup P_{m}\leq e^{-a.c}.\]
This allows us to apply the \emph{Lemma \ref{lem_continuity_L2} of
$L^{2}$-continuity} and obtain that for any $\varepsilon>0$, $\hat{P}_{m}$
decomposes as \[
\hat{P}_{m}=\hat{p}_{\varepsilon}+\hat{k}_{\varepsilon}\]
 with $\hat{k}_{\varepsilon}\in\Psi^{-\infty}$ a smoothing operator
and $\left\Vert \hat{p}_{\varepsilon}\right\Vert \leq e^{-a.c}+\varepsilon$.
Finally, we multiply on the left by $\hat{F}$ to obtain \[
\hat{Q}_{m}=\hat{F}\hat{p}_{\varepsilon}+\hat{F}\hat{k}_{\varepsilon}.\]
The second term is smoothing, hence compact, while the first one has
an operator norm bounded by $\left(e^{-ac}+\varepsilon\right)\left\Vert \hat{F}\right\Vert =Ce^{-ac}$,
with any $C>\left\Vert \hat{F}\right\Vert $ and the choice $\varepsilon=e^{-ac}\left(\frac{C-\left\Vert \hat{F}\right\Vert }{\left\Vert \hat{F}\right\Vert }\right)$.
We have shown the claimed spectral results for $\hat{Q}_{m}:L^{2}\left(M\right)\rightarrow L^{2}\left(M\right)$
and therefore for $\hat{F}:H^{m}\rightarrow H^{m}$.

\end{proof}
\normalsize

\vspace{0.cm}\begin{center}{\color{blue}\fbox{\color{black}\parbox{16cm}{
\begin{coro}

\label{cor:quasi-compact}Let $\varepsilon>0$ and let $m\in\mathcal{O}_{\varepsilon}$
be an order function as defined in Eq.(\ref{eq:def_O_eps}). If we
denote by $\pi$ the spectral projector associated to $\hat{F}:H^{m}\rightarrow H^{m}$
outside the disk of radius $\varepsilon$, and $\hat{K}\defi\hat{\pi}\hat{F}$,
$\hat{R}\defi\left(1-\hat{\pi}\right)\hat{F}$, then we have a spectral
decomposition \begin{equation}
\hat{F}=\hat{K}+\hat{R},\qquad\hat{K}\hat{R}=\hat{R}\hat{K}=0\label{eq:Decomp_F_K_R}\end{equation}
and
\begin{enumerate}
\item The spectral radius of $\hat{R}$ is smaller than $\varepsilon$.
\item $\hat{K}$ has finite rank. Its spectrum has generalized eigenvalues
$\lambda_{i}$ (counting multiplicity) called the \textbf{Ruelle resonances},
with $\varepsilon<\left|\lambda_{i}\right|$. The general Jordan decomposition
of $\hat{K}$ can be written \begin{equation}
\hat{K}=\sum_{i\geq0,\left|\lambda_{i}\right|>\varepsilon}\left(\lambda_{i}\sum_{j=1}^{d_{i}}v_{i,j}\otimes w_{i,j}+\sum_{j=1}^{d_{i}-1}v_{i,j}\otimes w_{i,j+1}\right)\label{eq:Jordan_f}\end{equation}
 with $d_{i}$ the dimension of the Jordan block associated to the
eigenvalue $\lambda_{i}$, with $v_{i,j}\in H^{m}$, $w_{i,j}\in H^{-m}$
($w_{i,j}$ is viewed as a linear form on $H^{m}$ with the duality
Eq.(\ref{eq:def_dualite})). They satisfy $w_{i,j}\left(v_{k,l}\right)=\delta_{ik}\delta_{jl}$. 
\item The distributions $v_{i,j},w_{i,j}$ and the corresponding eigenvalues
$\lambda_{i}$ do not depend on the choice of $m$, but are intrinsic
to the operator $\hat{F}$:\begin{equation}
v_{i,j}\in\left(\bigcap_{m\in\mathcal{O}_{\left|\lambda_{i}\right|}}H^{m}\right),\qquad w_{i,j}\in\left(\bigcap_{m\in\mathcal{O}_{\left|\lambda_{i}\right|}}H^{-m}\right).\label{eq:v_w_in_inter}\end{equation}
In other words, $v_{i,j}$ are smooth in every direction except in
the unstable direction which contains their wave front (\cite{taylor_tome2}
p.27):\[
WF\left(v_{i,j}\right)\subset E_{u}^{*}.\]
Similarly $w_{i,j}$ are smooth except in the stable direction which
contains their wave front:\[
WF\left(w_{i,j}\right)\subset E_{s}^{*}.\]
The resolvent $\left(z-\hat{F}\right)^{-1}$ has a meromorphic extension
from $C^{\infty}\left(M\right)$ to $\mathcal{D}'\left(M\right)$,
whose poles are the $\lambda_{i}$.
\end{enumerate}
\end{coro}
}}}\end{center}\vspace{0.cm}

\begin{proof}
\small 

Points 1 and 2 are immediate consequences of Theorem \ref{thm:Quasi-compact}.
The projector $\pi$ can be obtained from an integral of the resolvent
$\hat{R}\left(z\right)=\left(z-\hat{F}\right)^{-1}:H^{m}\rightarrow H^{m}$
on a circular contour of radius $\varepsilon$. 

We will prove now point 3 namely that the spectrum and eigen-distribution
do not depend on $m$. Let $\varepsilon>0$, and $m,m'\in\mathcal{O}_{\varepsilon}$
(defined in Eq.(\ref{eq:def_O_eps})), and suppose first that $m'\geq m$.
From Eq.(\ref{eq:Hm1_in_Hm2}), one has $H^{m'}\subset H^{m}$. Let
$\hat{F}_{m}$ (resp. $\hat{F}_{m'}$) denotes the restriction of
$\hat{F}$ to the distribution space $H^{m}$ (resp. $H^{m'}$). From
Theorem $\ref{thm:Quasi-compact}$, both $\hat{F}_{m}$ and $\hat{F}_{m'}$
are bounded operators and have essential spectrum radius less than
$\varepsilon$. For $\left|z\right|$ large enough, the resolvents
of $\hat{F}_{m}$ and $\hat{F}_{m'}$ are equal on $H^{m'}$ because
one can write\[
\hat{R}_{m'}\left(z\right)=\left(z-\hat{F}_{m'}\right)^{-1}=\frac{1}{z}\left(1+\sum_{n=1}^{\infty}\left(\frac{\hat{F}_{m'}}{z}\right)^{n}\right)=\hat{R}_{m}\left(z\right)\]
since $\hat{F}_{m'}=\hat{F}_{m}$ on $H^{m'}$ and the sum is convergent.
By meromorphic continuation, the resolvents also coincide for $\left|z\right|>\varepsilon$.
By explicit contour integral of the resolvent on a circle of radius
$\varepsilon$, one deduces that the corresponding finite rank operators
$\hat{K}_{m'}$ and $\hat{K}_{m}$ are equal on $H^{m'}$. But since
$H^{m'}$ is dense in $H^{m}$, one deduces that there are no eigenspaces
of $\hat{K}_{m}$ outside $H^{m'}$. Therefore the eigen-distributions
$v_{i,j}$ belongs to $H^{m'}$.

Now let $m,m''\in\mathcal{O}_{\varepsilon}$ be any two order functions
(with no mutual inclusion). From the explicit construction given in
the proof of Lemma \ref{lem:def_A_s_u}, one can find $m'\in\mathcal{O}_{\varepsilon}$,
such that $m'\geq m$ and $m'\geq m''$. Then the above argument shows
that $v_{i,j}\in H^{m'}\subset\left(H^{m}\cap H^{m''}\right)$.

Similar (but dual) arguments for the operator $\hat{F}^{*}:H^{-m}\rightarrow H^{-m}$
show that its eigenvectors $w_{i,j}$ are in $\left(H^{-m}\cap H^{-m''}\right)$\emph{.}
We have obtained Eq.(\ref{eq:v_w_in_inter}). 

Since $C^{\infty}\subset H^{m}$ for any $\varepsilon>0$, and any
$m\in\mathcal{O}_{\varepsilon}$, and $H^{m}\subset\mathcal{D}'\left(M\right)$,
we deduce that $\left(z-\hat{F}\right)^{-1}:C^{\infty}\left(M\right)\rightarrow\mathcal{D}'\left(M\right)$
admits a meromorphic extension on $\mathbb{C}\setminus\left\{ 0\right\} $.

\end{proof}
\normalsize

\section{Asymptotic expansion for dynamical correlation functions}

One usually obtains much information on $\hat{F}$ and the dynamical
system $f$ through the study of dynamical correlation functions.

\vspace{0.cm}\begin{center}{\color{red}\fbox{\color{black}\parbox{16cm}{
\begin{Def}

For $\psi_{1},\psi_{2}\in L^{2}\left(M\right)$ and $n\in\mathbb{Z}$,
define the \textbf{Lebesgue dynamical correlation function}\begin{equation}
C_{\psi_{2},\psi_{1}}^{Leb}\left(n\right)\defi\left(\psi_{2},\hat{F}^{n}\psi_{1}\right)=\int\overline{\psi}_{2}\left(x\right)\psi_{1}\left(f^{n}\left(x\right)\right)d\mu_{Leb}\label{eq:def_C_Psi1_PSi2}\end{equation}

\end{Def}
}}}\end{center}\vspace{0.cm}

\vspace{0.cm}\begin{center}{\color{red}\fbox{\color{black}\parbox{16cm}{
\begin{Def}

\label{def:Lebesgue_mixing}The diffeomorphism $f$ is called \textbf{Lebesgue-mixing}
if there exists an invariant measure%
\footnote{Notice that taking $\psi_{2}=1$, Eq.(\ref{eq:C_leb}) implies that
$\mu_{srb}$ is the limit of $f_{*}^{n}\left(\mu_{Leb}\right)$, for
$n\rightarrow\infty$.%
} $\mu_{srb}$ called the \textbf{Sinai-Ruelle-Bowen (SRB) measure}
such that for any \textbf{$\psi_{1},\psi_{2}\in C^{\infty}\left(M\right)$,\begin{equation}
C_{\psi_{2},\psi_{1}}^{Leb}\left(n\right)\underset{n\rightarrow\infty}{\longrightarrow}\left(\int\overline{\psi}_{2}d\mu_{Leb}\right)\left(\int\psi_{1}d\mu_{srb}\right)\label{eq:C_leb}\end{equation}
}

\end{Def}
}}}\end{center}\vspace{0.cm}

\vspace{0.cm}\begin{center}{\color{red}\fbox{\color{black}\parbox{16cm}{
\begin{Def}

For $\psi_{1},\psi_{2}\in L^{2}\left(M\right)$ and $n\in\mathbb{Z}$,
define the \textbf{SRB dynamical correlation function}\begin{equation}
C_{\psi_{2},\psi_{1}}^{srb}\left(n\right)\defi\int\overline{\psi}_{2}\left(x\right)\psi_{1}\left(f^{n}\left(x\right)\right)d\mu_{srb}\label{eq:def_C_srb}\end{equation}

\end{Def}
}}}\end{center}\vspace{0.cm}

We show in Theorem \ref{thm:SRB-mixing} that Lebesgue-mixing implies
\textbf{SRB-mixing}, i.e.\textbf{\begin{equation}
C_{\psi_{2},\psi_{1}}^{srb}\left(n\right)\underset{n\rightarrow\infty}{\longrightarrow}\left(\int\overline{\psi}_{2}d\mu_{srb}\right)\left(\int\psi_{1}d\mu_{srb}\right)\label{eq:C_srb}\end{equation}
}which is also the usual definition of mixing.

Let us mention the following conjecture (\cite{katok_hasselblatt}
p. 575, foot-note (2), or \cite{pesin_04} p. 7.)

\vspace{0.cm}\begin{center}{\color{red}\fbox{\color{black}\parbox{16cm}{
\begin{conj}

\label{con:mixing}A smooth Anosov diffeomorphism $f:M\rightarrow M$
on a connected compact manifold $M$ is Lebesgue-mixing.

\end{conj}
}}}\end{center}\vspace{0.cm}

In the particular case where $f$ preserves a smooth measure $dx$
(so $\hat{F}$ is unitary on $L^{2}\left(M\right)$), this has been
proved by Anosov in his PhD-thesis \cite{anosov_67} (see \cite{brin-02}
Theorem 6.3.1). In Section \ref{sec:Mixing-of-Anosov} we provide
a different proof entirely based on the semi-classical approach developed
in this paper.

We will assume Lebesgue-mixing in Theorem \ref{thm:SRB-mixing}.

Theorems \ref{thm:Ruelle-resonances} and \ref{thm:SRB-mixing} below
have been obtained before with various degrees of generality.

\subsection{The Lebesgue correlation function}

The Ruelle resonances $\lambda_{i}$ and associated distributions
$v_{i,j},w_{i,j}$ have been defined in Corollary \ref{cor:quasi-compact}.

\vspace{0.cm}\begin{center}{\color{blue}\fbox{\color{black}\parbox{16cm}{
\begin{theo}

\label{thm:Ruelle-resonances} For any $\psi_{1},\psi_{2}\in C^{\infty}\left(M\right)$,
$\varepsilon>0$ such that $\varepsilon\neq\left|\lambda_{i}\right|,\forall i$,
and $n\geq1$, one has\begin{equation}
C_{\psi_{2},\psi_{1}}^{Leb}\left(n\right)=\sum_{i\geq0,\left|\lambda_{i}\right|>\epsilon}\sum_{k=0}^{min\left(n,d_{i}-1\right)}C_{n}^{k}\lambda_{i}^{n-k}\sum_{j=1}^{d_{i}-k}v_{i,j}\left(\overline{\psi_{2}}\right)w_{i,j+k}\left(\psi_{1}\right)+\left\Vert \psi_{1}\right\Vert _{H^{m}}\left\Vert \psi_{2}\right\Vert _{H^{-m}}O_{\varepsilon}\left(\varepsilon^{n}\right).\label{eq:C_t_psi_1_psi2}\end{equation}
with any $m\in\mathcal{O}_{\varepsilon}$ (defined in Eq.(\ref{eq:def_O_eps}))
and $C_{n}^{k}:=\frac{n!}{\left(n-k\right)!k!}$. 

\end{theo}
}}}\end{center}\vspace{0.cm}

\paragraph{Remarks:}
\begin{itemize}
\item More generally Eq.(\ref{eq:C_t_psi_1_psi2}) still holds for $\psi_{1}\in H^{m}$
and $\psi_{2}\in H^{-m}$, with $m\in\mathcal{O}_{\varepsilon}$.
\item The right hand side of Eq.(\ref{eq:C_t_psi_1_psi2}) is complicated
by the possible presence of {}``Jordan blocks''. In the case where
the spectrum $\lambda_{i}$ is simple ($\lambda_{i}\neq\lambda_{j}$)
it reads more simply \[
C_{\psi_{2},\psi_{1}}^{Leb}\left(n\right)=\sum_{i\geq0,\left|\lambda_{i}\right|>\epsilon}\lambda_{i}^{n}v_{i}\left(\overline{\psi_{2}}\right)w_{i}\left(\psi_{1}\right)+O_{\varepsilon}\left(\varepsilon^{n}\right).\]

\end{itemize}
\begin{proof}
\small 

Theorem \ref{thm:Ruelle-resonances} is deduced from Corollary \ref{cor:quasi-compact}.
For any $\varepsilon>0$, let $m\in\mathcal{O}_{\varepsilon}$. For
any $n\geq0$ we have $\hat{F}^{n}=\hat{K}^{n}+\hat{R}^{n}$ and $\left\Vert \hat{R}^{n}\right\Vert _{H^{m}}=O_{\varepsilon}\left(\varepsilon^{n}\right)$.
If $\psi_{1}\in H^{m}$, $\psi_{2}\in H^{-m}$ then we use Eq.(\ref{eq:Inegalite})
to write \begin{eqnarray}
C_{\psi_{2},\psi_{1}}^{Leb}\left(n\right) & = & \left(\psi_{2},\hat{F}^{n}\psi_{1}\right)\label{eq:CC}\\
 & = & \left(\psi_{2},\hat{K}^{n}\psi_{1}\right)+\left(\psi_{2},\hat{R}_{a}^{n}\psi_{1}\right)\nonumber \\
 & = & \left(\psi_{2},\hat{K}^{n}\psi_{1}\right)+\left\Vert \psi_{2}\right\Vert _{H^{m}}\left\Vert \psi_{1}\right\Vert _{H^{-m}}O_{\varepsilon}\left(\varepsilon^{n}\right)\nonumber \end{eqnarray}
Using the Jordan Block decomposition of $\hat{K}$, Eq.(\ref{eq:Jordan_f}),
and Eq.(\ref{eq:def_dualite}), we have \begin{eqnarray}
\left(\psi_{2},\hat{K}^{n}\psi_{1}\right) & = & \sum_{i\geq0,\left|\lambda_{i}\right|>\varepsilon}\sum_{k=0}^{min\left(n,d_{i}-1\right)}C_{n}^{k}\lambda_{i}^{n-k}\sum_{j=1}^{d_{i}-k}\left(\psi_{2},v_{i,j}\right)w_{i,j+k}\left(\psi_{1}\right),\label{eq:Jordan_f2}\\
 & = & \sum_{i\geq0,\left|\lambda_{i}\right|>\varepsilon}\sum_{k=0}^{min\left(n,d_{i}-1\right)}C_{n}^{k}\lambda_{i}^{n-k}\sum_{j=1}^{d_{i}-k}v_{i,j}\left(\overline{\psi_{2}}\right)w_{i,j+k}\left(\psi_{1}\right)\end{eqnarray}
We have obtained Eq.(\ref{eq:C_t_psi_1_psi2}).

\end{proof}
\normalsize

\vspace{0.cm}\begin{center}{\color{blue}\fbox{\color{black}\parbox{16cm}{
\begin{coro}

\label{cor:spectral_radius_1}For any $i\geq0$, $\left|\lambda_{i}\right|\leq1$.
Therefore the spectral radius of the operator $\hat{F}:H^{m}\rightarrow H^{m}$
is one. If an eigenvalue is on the unit circle, $\left|\lambda_{i}\right|=1$,
then $d_{i}=1$, i.e. it has no Jordan block.

\end{coro}
}}}\end{center}\vspace{0.cm}

\begin{proof}
\small 

Since $\hat{F}\varphi=\varphi\circ f$, it is clear that $\lambda_{0}=1$
is an eigenvalue for the constant function. For all $n$, and $\psi_{1},\psi_{2}\in C^{\infty}\left(M\right)$,
one has $\left(\psi_{2},\hat{F}^{n}\psi_{1}\right)=\int\overline{\psi}_{2}\left(x\right)\psi_{1}\left(f^{n}\left(x\right)\right)d\mu_{Leb}$
hence \begin{equation}
\left|C_{\psi_{1},\psi_{2}}^{Leb}\left(n\right)\right|=\left|\left(\psi_{2},\hat{F}^{n}\psi_{1}\right)\right|\leq\left|\psi_{2}\right|_{C^{0}}\left|\psi_{1}\right|_{C^{0}}\mbox{Vol}\left(M\right)\label{eq:C_bounded}\end{equation}
is bounded uniformly with respect to $n$. 

Suppose that $\lambda_{i}>1$. Since $C^{\infty}\left(M\right)$ is
dense in $H^{m}$ and $H^{-m}$, there exists $\psi_{1},\psi_{2}\in C^{\infty}\left(M\right)$
such that $v_{i,1}\left(\overline{\psi_{2}}\right)\neq0$ and $w_{i,1}\left(\psi_{1}\right)\neq0$.
Then $\lambda_{i}^{n}v_{i,1}\left(\overline{\psi_{2}}\right)w_{i,1}\left(\psi_{1}\right)$
would diverge for $n\rightarrow\infty$, and Eq.(\ref{eq:C_t_psi_1_psi2})
implies that $C_{\psi_{2},\psi_{1}}^{Leb}\left(n\right)$ would diverge
also, in contradiction with Eq.(\ref{eq:C_bounded}).

Similarly, suppose that $\left|\lambda_{i}\right|=1$, but $d_{i}\geq2$.
There exists $\psi_{1},\psi_{2}\in C^{\infty}\left(M\right)$ such
that $v_{i,j}\left(\overline{\psi_{2}}\right)\neq0$ and $w_{i,j}\left(\psi_{1}\right)\neq0$.
Then the term $k=d_{i}-1$ in Eq.(\ref{eq:C_t_psi_1_psi2}) which
contains $C_{n}^{k}$ diverges as $n^{d_{i}-1}$ for $n\rightarrow\infty$.
Eq.(\ref{eq:C_t_psi_1_psi2}) implies that $C_{\psi_{2},\psi_{1}}^{Leb}\left(n\right)$
would diverge also, in contradiction with Eq.(\ref{eq:C_bounded}).

\end{proof}
\normalsize

\vspace{0.cm}\begin{center}{\color{blue}\fbox{\color{black}\parbox{16cm}{
\begin{coro}

\label{cor:mixing}The following two propositions are equivalent:
\begin{enumerate}
\item $f$ is Lebesgue-mixing. 
\item $\lambda_{0}=1$ is simple, $v_{0}=Leb$ is the Lebesgue measure and
$w_{0}=\mu_{srb}$ the SRB measure. The other eigenvalues satisfy
$\left|\lambda_{i}\right|<1$, $i\geq1$.
\end{enumerate}
\end{coro}
}}}\end{center}\vspace{0.cm}

Therefore:\[
C_{\psi_{2},\psi_{1}}^{Leb}\left(n\right)\underset{n\rightarrow\infty}{\longrightarrow}v_{0}\left(\overline{\psi_{2}}\right)w_{0}\left(\psi_{1}\right)\]

\paragraph{Remarks:}
\begin{itemize}
\item It turns out that the SRB correlation function $C_{\psi_{2},\psi_{1}}^{srb}\left(n\right)$
admits an asymptotic expansion similar to Eq.(\ref{eq:C_t_psi_1_psi2}),
see Theorem \ref{thm:SRB-mixing} below.
\item Without the Lebesgue-mixing assumption and if Conjecture \ref{con:mixing}
is wrong, there may be a finite number of eigenvalues on the unit
circle. 
\end{itemize}
\begin{proof}
\small 

The Lebesgue-mixing assumption implies that $C_{\psi_{1},\psi_{2}}^{Leb}\left(n\right)$
converges for $n\rightarrow\infty$. The constant function $v_{0}=1$
is obviously an eigenfunction of $\hat{F}$ with eigenvalue $\lambda_{0}=1$.
There are no other eigenvalues on the unit circle otherwise from Eq.(\ref{eq:C_t_psi_1_psi2}),
$C_{\psi_{1},\psi_{2}}^{Leb}\left(n\right)$ would not converge for
$n\rightarrow\infty$. We obtain $C_{\psi_{1},\psi_{2}}^{Leb}\left(n\right)=\left(\psi_{2},\hat{F}^{n}\psi_{1}\right)\underset{n\rightarrow\infty}{\rightarrow}v_{0}\left(\overline{\psi_{2}}\right)w_{0}\left(\psi_{1}\right)$
with $w_{0}=\mu_{srb}$ from Definition \ref{def:Lebesgue_mixing}.
But Eq.(\ref{eq:C_bounded}) also implies that $\left|w_{0}\left(\psi_{1}\right)\right|\leq C\left|\psi_{1}\right|_{C^{0}}$.
Therefore $w_{0}$ is distribution of order $0$, hence defines a
measure.

\end{proof}
\normalsize

\subsection{\label{sub:SRB-correlation-function}The SRB correlation function}

We have shown above that $\mu_{srb}=w_{0}\in H^{-m}$ for any $m\in\mathcal{O}_{\varepsilon}$
and $\varepsilon<1$. Notice that from Eq.(\ref{eq:f-phi}), we have
$w_{0}\psi_{2}\in H^{-m}$ for any $\psi_{2}\in C^{\infty}\left(M\right)$.
Since $v_{i,j}\in H^{m}$ then Eq.(\ref{eq:def_dualite}) implies
that $v_{i,j}\left(w_{0}\psi_{2}\right)=\left(v_{ij},w_{0}\psi_{2}\right)$
makes sense.

\vspace{0.cm}\begin{center}{\color{blue}\fbox{\color{black}\parbox{16cm}{
\begin{theo}

\label{thm:SRB-mixing}Assume that $f$ is Lebesgue-mixing. If $\psi_{1},\psi_{2}\in C^{\infty}\left(M\right)$
then the asymptotic behavior of the SRB correlation function Eq.(\ref{eq:C_srb})
is given by: \begin{eqnarray}
C_{\psi_{2},\psi_{1}}^{srb}\left(n\right) & = & w_{0}\left(\overline{\psi_{2}}\right)w_{0}\left(\psi_{1}\right)\label{eq:C_t_srb}\\
 &  & +\sum_{i\geq0,\left|\lambda_{i}\right|>\epsilon}\sum_{k=0}^{min\left(n,d_{i}-1\right)}C_{n}^{k}\lambda_{i}^{n-k}\sum_{j=1}^{d_{i}-k}v_{i,j}\left(\overline{w_{0}\psi_{2}}\right)w_{i,j}\left(\psi_{1}\right)+O_{\varepsilon}\left(\varepsilon^{n}\right).\nonumber \end{eqnarray}
In particular $f$ is SRB-mixing :\[
C_{\psi_{2},\psi_{1}}^{srb}\left(n\right)\underset{n\rightarrow\infty}{\longrightarrow}w_{0}\left(\overline{\psi_{2}}\right)w_{0}\left(\psi_{1}\right)=\left(\int\overline{\psi}_{2}d\mu_{srb}\right)\left(\int\psi_{1}d\mu_{srb}\right)\]
 and the convergence is exponentially fast.

\end{theo}
}}}\end{center}\vspace{0.cm}

\begin{proof}
\small 

Using Eq.(\ref{eq:def_dualite}) and Eq.(\ref{eq:relation}), we start
with an equivalent expression for the SRB correlation function:\[
C_{\psi_{2},\psi_{1}}^{srb}\left(n\right)=\int\overline{\psi}_{2}\left(x\right)\psi_{1}\left(f^{n}\left(x\right)\right)d\mu_{srb}=\left(w_{0},\overline{\psi}_{2}\hat{F}^{n}\psi_{1}\right)=\left(w_{0}\psi_{2},\hat{F}^{n}\psi_{1}\right)\]
Then as in Eq.(\ref{eq:CC}), we use the decomposition Eq.(\ref{eq:Jordan_f})
and deduce Eq.(\ref{eq:C_t_srb}), since $v_{0}\left(\overline{w_{0}\psi_{2}}\right)w_{0}\left(\psi_{1}\right)=w_{0}\left(\overline{\psi_{2}}\right)w_{0}\left(\psi_{1}\right)$.

\end{proof}
\normalsize

\section{\label{sec:Mixing-of-Anosov}Mixing of Anosov maps preserving a smooth
measure}

In the particular case where $f$ preserves a smooth measure $dx$,
ergodicity of $f$ and therefore mixing, has been proved by Anosov
in his PhD thesis \cite{anosov_67} (see \cite{brin-02} Theorem 6.3.1).
In this section we provide a different proof entirely based on the
semi-classical approach developed in this paper.

\vspace{0.cm}\begin{center}{\color{blue}\fbox{\color{black}\parbox{16cm}{
\begin{theo}

\label{pro:mixing}Suppose that $f$ preserves a smooth measure $dx$.
Then on the unit circle, there is no Ruelle resonance, except $1$
with multiplicity one (equivalently $f$ is Lebesgue-mixing from Corollary
\ref{cor:mixing}).

\end{theo}
}}}\end{center}\vspace{0.cm}

\begin{proof}
\small 

\emph{of Theorem \ref{pro:mixing}.} From Corollary \ref{cor:quasi-compact}
and Corollary \ref{cor:spectral_radius_1} an eigenvalue $\lambda=e^{i\theta}$
on the unit circle would have no Jordan Block and would correspond
to an eigen-vector $u\in H^{m}$, $\hat{F}u=e^{i\theta}u$. Lemma
\ref{erg1} and Lemma \ref{pro:mixing_C1} below imply that $u$ is
a constant function and $\lambda=1$.

\end{proof}
\normalsize

The following Lemma contains the global aspect of the problem.

\vspace{0.cm}\begin{center}{\color{blue}\fbox{\color{black}\parbox{16cm}{
\begin{lem}

\label{pro:mixing_C1}If %
\footnote{Let us remark that the hypothesis $C^{1}\left(M\right)$ is important.
The result is not true in the space of measures, since there are many
periodic orbits for $f$, and for each periodic orbit of period $N$,
there exists a Dirac measure $u$ supported on it such that $\hat{F}u=e^{i\theta}u$
with $e^{i\theta N}=1$.%
} $u\in C^{1}\left(M\right)$ and $\hat{F}u=\lambda u$, with $\left|\lambda\right|=1$
then $u$ is a constant function, and $\lambda=1$.

\end{lem}
}}}\end{center}\vspace{0.cm}

\begin{proof}
\small 

Let us assume that $u\in C^{1}\left(M\right)$, with $\hat{F}u=\lambda u$,
$\left|\lambda\right|=1$. Let $u_{n}:=\hat{F}^{n}u=u\circ f^{n}$.
One has $u_{n}=\lambda^{n}u$, therefore $\left|du_{n}\right|_{\infty}=\left|du\right|_{\infty}<\infty$
is bounded uniformly with respect to $n$, since $M$ is compact.
On the other hand $\left(du_{n}\right)_{f^{-n}\left(x\right)}=\left(Df^{n}\right)_{f^{-n}x}^{\mathbf{t}}\left(du\right)_{x}$.
Suppose that there exists $x\in M$ such that $du_{x}\neq0$. If $du_{x}\notin E_{s}^{*}\left(x\right)$
(stable direction) then $\left|\left(du_{n}\right)_{f^{-n}\left(x\right)}\right|$
diverge when $n\rightarrow+\infty$. If $du_{x}\in E_{s}^{*}\left(x\right)$
then $\left|\left(du_{n}\right)_{f^{-n}\left(x\right)}\right|$ diverge
when $n\rightarrow-\infty$. This contradicts $\left|du_{n}\right|_{\infty}<\infty$,
therefore $du=0$. $M$ is connected therefore $u$ is constant.

\end{proof}
\normalsize

\vspace{0.cm}\begin{center}{\color{blue}\fbox{\color{black}\parbox{16cm}{
\begin{lem}

\label{erg1}Assume there exists a positive smooth density $dx$ on
$M$ which is invariant under the map $f$. Let $u\in H^{m}$, $\hat{F}u=e^{i\theta}u$.
Then $u\in C^{\infty}\left(M\right)$.

\end{lem}
}}}\end{center}\vspace{0.cm}

\begin{proof}
\small 

We shall make use of some $h$-pseudo\-differential calculus%
\footnote{In particular $\xi$ is quantized into the operator $hD_{x}=-ih\partial/\partial x$
while for ordinary PDO, $\xi$ is quantized into $D_{x}=-i\partial/\partial x$.%
}. Assume there exists a positive smooth density $dx$ on $M$ which
is invariant under the map $f$. Then $\hat{F}:L^{2}(M,dx)\to L^{2}(M,dx)$
is unitary. The idea of the proof is to use Corollary \ref{cor:quasi-compact}
which states that $u$ is $C^{\infty}$ in every direction except
the unstable direction, and use the unitary of $\hat{F}$, to deduce
that $u$ is $C^{\infty}$ also in the unstable direction (by propagation).

It is easy to see that there exist symbols \[
0\le B(x,\xi),C(x,\xi)\in S^{0}(T^{*}M),\]
 such that \begin{equation}
1=B^{2}+C^{2},\label{erg.0}\end{equation}
and $B(x,\xi)=1$ on the set \begin{equation}
\Vert\xi_{u}\Vert^{2}\le(1-\delta)(1+\Vert\xi_{s}\Vert^{2})\label{erg.1}\end{equation}
 with support in \begin{equation}
\Vert\xi_{u}\Vert^{2}\le(1+\delta)(1+\Vert\xi_{s}\Vert^{2}),\label{erg.2}\end{equation}
 where we choose $\delta>0$ sufficiently small. See Figure \ref{fig:B-D}.

\begin{figure}
\begin{centering}
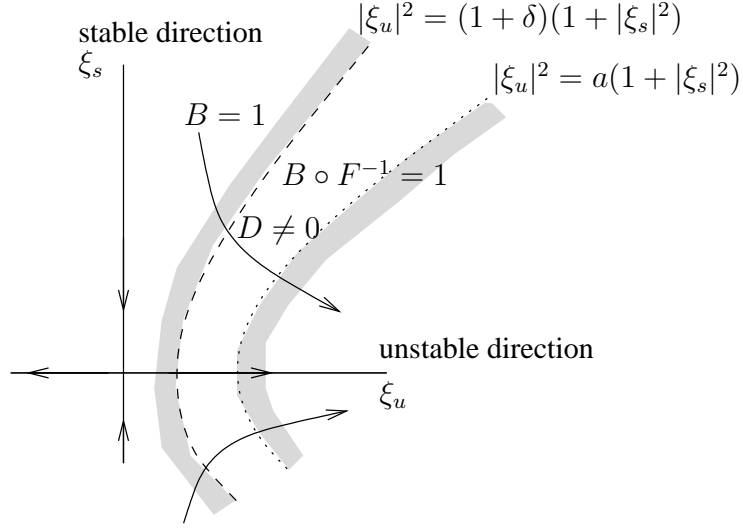
\par\end{centering}

\caption{\label{fig:B-D}A distribution $u\in H^{m}$ is regular in the stable
direction. If $\hat{F}u=e^{i\theta}u$, the idea of the proof of Lemma
\ref{erg.1} is to propagate this regularity under the map $F$ towards
the unstable direction. For that purpose, we propagate the symbol
$B$ and establish in (\ref{erg.9}) that $u$ is semi-classically
negligible in a zone where $D\neq0$.}

\end{figure}

Then $B\circ F^{-1}\in S^{0}$ is equal to 1 on a set \begin{equation}
\Vert\xi_{u}\Vert^{2}\le a(1+\Vert\xi_{s}\Vert^{2})\label{erg.3}\end{equation}
 and has its support in a set \begin{equation}
\Vert\xi_{u}\Vert^{2}\le b(1+\Vert\xi_{s}\Vert^{2}),\label{erg.4}\end{equation}
 where $1<a<b$ are independent of $\delta$ when $\delta>0$ is small
enough. Now we can construct corresponding $h$-pseudodifferential
operators $\hat{B},\hat{C}$ such that \begin{equation}
1=\hat{B}^{2}+\hat{C}^{2}+K,\label{erg.5}\end{equation}
 where $K$ is negligible in the sense that \begin{equation}
K={\cal O}(h^{N}):H^{-N}\to H^{N},\forall N\in\mathbb{N},\label{erg.6}\end{equation}
 and such that the symbol of $\hat{B}$ is equal to $B$ modulo $hS^{-1}$,
and modulo $h^{\infty}S^{-\infty}$ it is equal to 1 on the set (\ref{erg.1})
and has its support in the set (\ref{erg.2}). It follows from Egorov
Theorem that $\hat{F}\hat{B}\hat{F}^{-1}=\hat{F}\hat{B}\hat{F}^{*}$
has the corresponding properties with the sets (\ref{erg.1}), (\ref{erg.2})
replaced by (\ref{erg.3}), (\ref{erg.4}).

We can find a self-adjoint $h$-pseudodifferential operator $\hat{D}$
with symbol of class $S^{0}$, such that \begin{equation}
(\hat{F}\hat{B}\hat{F}^{*})^{2}-\hat{B}^{2}=\hat{D}^{2}+L,\label{erg.7}\end{equation}
 where $L$ is negligible as in (\ref{erg.6}). In fact, in the region
(\ref{erg.1}) we can take $\hat{D}=0$ and when we further approach
the unstable directions we first have $\hat{F}\hat{B}\hat{F}^{*}=1$
micro-locally, so that the left hand side in (\ref{erg.7}) is $\equiv1-\hat{B}^{2}\equiv\hat{C}^{2}$,
so that we can take $\hat{D}=\hat{C}$. Even closer to the unstable
directions, we get outside the support of $\hat{B}$ and we can take
$\hat{D}=\hat{F}\hat{B}\hat{F}^{*}.$

Now, let $u\in H^{m}$ be as in the proposition and write \[
\hat{B}u=e^{i\theta}\hat{B}\hat{F}^{-1}u.\]
 Thanks to the properties of $\hat{B}$ and $m$, this quantity belongs
to $L^{2}$, and using the unitarity of $\hat{F}$, we get \[
\Vert\hat{B}u\Vert^{2}=\Vert\hat{F}\hat{B}\hat{F}^{-1}u\Vert^{2}.\]
 Combining this with (\ref{erg.7}), we get \begin{equation}
0=\Vert\hat{F}\hat{B}\hat{F}^{-1}u\Vert^{2}-\Vert\hat{B}u\Vert^{2}=\Vert\hat{D}u\Vert^{2}+(Lu|u),\label{erg.8}\end{equation}
 and since $L$ is negligible, \begin{equation}
\Vert\hat{D}u\Vert={\cal O}(h^{\infty}).\label{erg.9}\end{equation}
 Since $\hat{D}$ is semi-classically elliptic in the region \[
(1+\delta)(1+\Vert\xi_{s}\Vert^{2})\le\Vert\xi_{u}\Vert^{2}\le a(1+\Vert\xi_{s}\Vert^{2}),\]
 we see that $u$ is micro-locally ${\cal O}(h^{\infty})$ in the
region (replacing $\xi\rightarrow h\xi$) \[
(1+\delta)(\frac{1}{h^{2}}+\Vert\xi_{s}\Vert^{2})\le\Vert\xi_{u}\Vert^{2}\le a(\frac{1}{h^{2}}+\Vert\xi_{s}\Vert^{2}),\]
 and letting $h\to0$, we see that $u$ has no wave-front set in a
conical neighborhood of $E_{u}^{*}$. Since we already know that $\mathrm{WF}(u)\subset E_{u}^{*}$,
we conclude that $u\in C^{\infty}$, and this ends the proof of Lemma
\ref{erg1}.

\end{proof}
\normalsize

\section{Truncation and numerical calculation of the resonance spectrum}

Let $\chi:\mathbb{R}^{+}\rightarrow\mathbb{R}^{+}$ be a $C^{\infty}$
function  such that $\chi(x)=1$, if $x\leq1$, and $\chi\left(x\right)=0$,
if $x\geq2$. For $r>0$, let the function $\chi_{r}$ on $T^{*}M$
be defined by $\chi_{r}\left(x,\xi\right)=\chi\left(\left|\xi\right|/r\right)$.
Let the \textbf{truncation operator} be:\[
\hat{\chi_{r}}\defi Op\left(\chi_{r}\right)\]
Notice that $\hat{\chi_{r}}$ is a smoothing operator which truncates
large components in $\xi$. In \cite{fred-RP-06} section 2.1.4 and
references therein, we interpret $\hat{\chi}_{r}$ as a {}``noisy
operator'' with a noise of amplitude $1/r$.

\vspace{0.cm}\begin{center}{\color{blue}\fbox{\color{black}\parbox{16cm}{
\begin{theo}

\label{thm:truncation}$\left(\hat{F}\hat{\chi_{r}}\right)$ is a
smoothing operator. For any $\varepsilon>0$ the spectrum of $\left(\hat{F}\hat{\chi_{r}}\right)$
in $L^{2}\left(M\right)$ outside the disk of radius $\varepsilon$,
converges for $r\rightarrow\infty$, towards the spectrum of Ruelle
resonances $\left(\lambda_{i}\right)_{i}$, counting multiplicities.
The eigenspaces converge towards the eigen-distributions.

\end{theo}
}}}\end{center}\vspace{0.cm}

\begin{figure}
\begin{centering}
\scalebox{1.5}[1.5]{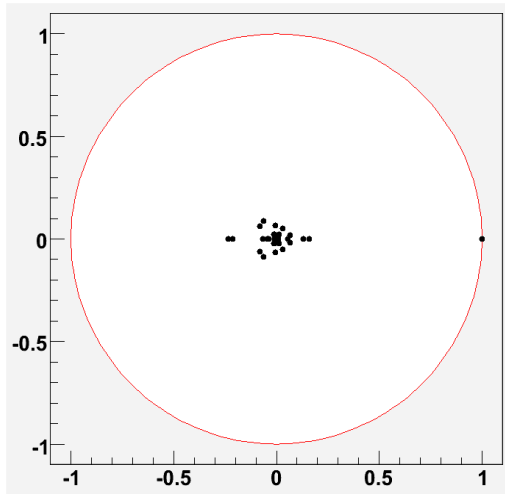}
\par\end{centering}

\caption{\label{fig:Ruelle-Resonances-of}Ruelle resonances $\lambda_{i}$
obtained numerically for the model Eq.(\ref{eq:example_on_T2}), with
$\varepsilon=0.5$. }

\end{figure}

\paragraph*{Remarks:}
\begin{enumerate}
\item Theorem \ref{thm:truncation} gives a practical way to compute numerically
the resonance spectrum: one expresses the operator $\hat{F}$ in a
discrete basis of $L^{2}\left(M\right)$, truncates it smoothly (according
to the operator $\hat{\chi}_{r}$), and diagonalizes the resulting
matrix numerically. See Figure \ref{fig:Ruelle-Resonances-of}.
\item Theorem \ref{thm:truncation} gives also a simple way to establish
a relation between Ruelle resonances $\lambda_{i}$ and the periodic
points of the map $f$, via dynamical zeta functions, see e.g. \cite{liverani_05,baladi_06}.
On one hand the Atiyah-Bott fixed point formula (\cite{atiyah_67}
corollary 5.4 p.393), gives for any $n\geq1$, \[
\mathbf{Tr}\left(\left(\hat{F}\hat{\chi_{r}}\right)^{n}\right)\underset{r\rightarrow\infty}{\longrightarrow}\sum_{x\in Fix\left(f^{n}\right)}\frac{1}{\left|\det\left(1-D_{x}f^{n}\right)\right|}\]
and on the other hand, the zeros of the dynamical zeta function \[
d\left(z\right)=\exp\left(-\sum_{n\geq1}\frac{z^{n}}{n}\mathbf{Tr}\left(\left(\hat{F}\hat{\chi_{r}}\right)^{n}\right)\right)=\mathbf{det}\left(1-z\left(\hat{F}\hat{\chi_{r}}\right)\right)\]
 converge towards $\left(1/\lambda_{i}\right)_{i}$.
\end{enumerate}
\begin{proof}
\small 

Let $\varepsilon>0$, and $m\in\mathcal{O}_{\varepsilon/2}$. From
Eq.(\ref{eq:F_rk}), the operator $\hat{F}:H^{m}\rightarrow H^{m}$
can be written $\hat{F}=\hat{r}+\hat{k}$, with $\left\Vert \hat{r}\right\Vert \leq\varepsilon/2$,
and $\hat{k}$ compact. In $H^{m}$, the operator $\hat{\chi}_{r}\underset{r\rightarrow\infty}{\longrightarrow}Id$
converges strongly, and $\left\Vert \hat{\chi_{r}}\right\Vert \leq C_{r}\underset{r\rightarrow\infty}{\longrightarrow}1$,
in particular $\left\Vert \hat{\chi_{r}}\right\Vert \leq2$ for $r$
large enough. For $\left|z\right|>\varepsilon$ write\[
\hat{F}-z=\hat{r}-z+\hat{k}=\left(\hat{r}-z\right)\left(1+\left(\hat{r}-z\right)^{-1}\hat{k}\right)\]
\[
\left(\hat{F}-z\right)^{-1}=\left(1+\left(\hat{r}-z\right)^{-1}\hat{k}\right)^{-1}\left(\hat{r}-z\right)^{-1}\]
when the first factor on the right is well defined. Similarly\[
\left(\hat{F}\hat{\chi}_{r}-z\right)^{-1}=\left(1+\left(\hat{r}\hat{\chi}_{r}-z\right)^{-1}\hat{k}\hat{\chi}_{r}\right)^{-1}\left(\hat{r}\hat{\chi}_{r}-z\right)^{-1}.\]
Let $z\in\mathbb{C}\setminus\sigma\left(\hat{F}\right)$ and $\left|z\right|>\varepsilon$.
Then $\left(\hat{r}\hat{\chi}_{r}-z\right)^{-1}$ is well defined
since $\left\Vert \hat{r}\hat{\chi}_{r}\right\Vert \leq\varepsilon$.
Moreover $\left(\hat{r}\hat{\chi}_{r}-z\right)^{-1}\underset{r\rightarrow\infty}{\longrightarrow}\left(\hat{r}-z\right)^{-1}$
strongly and $\left(\hat{r}\hat{\chi}_{r}-z\right)^{-1}\hat{k}\hat{\chi}_{r}\underset{r\rightarrow\infty}{\longrightarrow}\left(\hat{r}-z\right)^{-1}\hat{k}$
in norm since $\hat{k}$ is compact. Therefore $\left(\hat{F}\hat{\chi}_{r}-z\right)^{-1}\underset{r\rightarrow\infty}{\longrightarrow}\left(\hat{F}-z\right)^{-1}$
converges strongly. Now the finite rank operator $\hat{K}$ in the
spectral decomposition Eq.(\ref{eq:Decomp_F_K_R}) can be obtained
by a contour integral of the resolvent:\[
\hat{\pi}=\frac{1}{2\pi i}\oint_{\left|z\right|=\varepsilon}\left(z-\hat{F}\right)^{-1}dz,\qquad\hat{K}=\hat{\pi}\hat{F}\]
Similarly the spectrum of $\left(\hat{F}\hat{\chi}_{r}\right)$ outside
the disk of radius $\varepsilon$ is the spectrum of the finite rank
operator $\hat{K}_{r}$ given by: \[
\hat{\pi}_{r}\defi\frac{1}{2\pi i}\oint_{\left|z\right|=\varepsilon}\left(z-\hat{F}\hat{\chi}_{r}\right)^{-1}dz,\qquad\hat{K}_{r}=\hat{\pi}_{r}\left(\hat{F}\hat{\chi}_{r}\right).\]

The operator $\hat{K}_{r}\underset{r\rightarrow\infty}{\longrightarrow}\hat{K}$
converges strongly and therefore in norm since it has finite rank
(see Th. 9.19 p.98 in \cite{hislop_95}). We deduce Theorem \ref{thm:truncation}.

\end{proof}
\normalsize

\section{Conclusion and perspectives}

In this paper we have proposed a semi-classical approach for spectral
properties of Anosov dynamical systems. We discuss now some possible
perspectives for this work. 

First, we have treated the case of hyperbolic diffeomorphisms $f$.
The case of expansive maps which are not invertible is more simple,
and the method we propose works equally well, but need some adaptations.
For example the transfer operator $\hat{F}$ has the same definition
Eq.(\ref{eq:def_F^}), but the associated canonical map $F$ on $T^{*}M$
is now multivalued (its graph on $T^{*}M\times T^{*}M$ is well defined).

The case of partially hyperbolic systems (and in particular hyperbolic
flows) where there is a neutral direction is very interesting, and
there are important recent results concerning their spectral properties
\cite{dolgopyat_98,liverani_contact_04}. For the same reasons explained
in the introduction, we think that a semi-classical approach is natural
and hopefully fruitful for these systems too.

Finally let us mention the open question mentioned in the conjecture
\ref{con:mixing}. One can wonder if a semi-classical approach similar
to Section \ref{sec:Mixing-of-Anosov} could help towards the resolution
of this.

\appendix

\section{\label{Appendix:PDO's-with-variable}Pseudodifferential operators
with variable order}

\subsection{Preliminary remarks}

\subsubsection{Semi-classical analysis}

In this appendix, we provide a self-contained series of analytic tools
for studying pseudodifferential operators with slightly more general
classes of symbols than usual ones, namely symbols with variable order.
All the results we give come from the standard semi-classical analysis
but for our special symbol classes, and are given mainly without proof.
We refer to \cite{melrose_cours,grigis_sjostrand,taylor_tome2} for
the standard results in semi-classical analysis. Note that the idea
of using symbols with variable order is not new. See for example \cite{malgrange_59,unterberger,leopold_91}.

Semi-classical analysis is a rich theory which gives a sense to pseudodifferential
operators $\hat{P}$ of the form \begin{equation}
\varphi\rightarrow\hat{P}\left(\varphi\right)\left(x\right)=\int_{\mathbb{R}^{2d}}e^{i\xi\left(x-y\right)}P\left(x,\xi\right)\varphi\left(y\right)dyd\xi\label{eq_pdo}\end{equation}
where $P\left(x,\xi\right)$ is a smooth function called the \textbf{symbol}
of $\hat{P}$ and belonging to some appropriate class of functions
satisfying certain regularity conditions at infinity. These classes
yield to a powerful \textbf{symbol calculus}, i.e., a tool for extracting
information in an asymptotic way about the operator $\hat{P}$ by
means of its symbol. The main results are basically the following
:
\begin{itemize}
\item Composition of PDO's. Given two PDO's $\hat{P}$ and $\hat{Q}$, the
product $\hat{A}=\hat{P}\hat{Q}$ is also a PDO and its symbol is
given to leading order in $\xi$ by the product of the symbols $A=PQ$.
\item Sobolev continuity. First defined on $C^{\infty}\left(\mathbb{R}^{n}\right)$,
pseudodifferential operators are shown to be continuous between certain
Sobolev spaces. Moreover, the corresponding operator norm is estimated
by some norms of the derivatives of the symbol. 
\item Ellipticity, parametrix. A condition called \textbf{ellipticity} imposed
on a symbol is enough to insure that the corresponding operator is
invertible up to a regularizing operator. The {}``almost-inverse''
is called a parametrix. 
\end{itemize}

\subsubsection{Symbols with constant order }

The typical class of symbols one considers is the set $S_{\rho,\delta}^{m}\subset C^{\infty}\left(\mathbb{R}^{2n}\right)$
of smooth functions $P\left(x,\xi\right)$ which satisfy the following
estimates. For any compact subset $K\subset\mathbb{R}^{n}$ and any
multi-index $\alpha,\beta\in\mathbb{N}^{n}$, there is a constant
$C_{K,\alpha,\beta}$ such that \begin{equation}
\left|\partial_{\xi}^{\alpha}\partial_{x}^{\beta}P\left(x,\xi\right)\right|\leq C_{K,\alpha,\beta}\left\langle \xi\right\rangle ^{m-\rho\left|\alpha\right|+\delta\left|\beta\right|}\label{eq_symbol}\end{equation}
for any $\left(x,\xi\right)\in K\times\mathbb{R}^{n}$. Here $\left\langle \xi\right\rangle $
means $\sqrt{1+\left|\xi\right|^{2}}$ and we use the standard multi-indices
notation $\partial_{x}^{\alpha}f=\frac{\partial^{\alpha_{i}}f}{\partial x_{i}^{\alpha_{i}}}...\frac{\partial^{\alpha_{n}}f}{\partial x_{n}^{\alpha_{n}}}$.
The number $m\in\mathbb{R}$ is called the (\textbf{constant}) \textbf{order}
of the symbol. These classes of symbols, introduced by Hörmander \cite{hormander_3},
are quite general and allow one nevertheless to develop a symbol calculus,
provided the constants $\rho,\delta$ fulfill certain conditions.
The corresponding class of PDO's (Formula \ref{eq_pdo}) is denoted
by $\Psi_{\rho,\delta}^{m}$. The operator $\hat{P}$, denoted sometimes
also by $P\left(x,D\right)$, is the (\textbf{left}) \textbf{quantization}
of $P$.

Nevertheless, as explained in the introduction, we need to consider
symbols with an order $m$ which is no longer constant but depends
on the variables $\left(x,\xi\right)$. It turns out that these classes
are contained in some Hörmander classes (with order equal to $\limsup m$)
but one needs to keep track of the fact that the order is variable
in order to develop a more general notion of ellipticity, for symbols
which would not be elliptic in the usual sense. This will be explained
in the rest of this appendix which is devoted to the symbol calculus
for symbols with variable order. But before, we need one additional
remark about the theory of PDO's on manifolds.

\subsubsection{Pseudodifferential operators on manifold}

The usual way of defining PDO's on a para-compact Hausdorff manifold
$M$ is to make a partition of the unity of $M$ and use in each chart
the semi-classical analysis on $\mathbb{R}^{n}$. Defined in this
way, the symbol of an operator depends unfortunately on the charts.
But it turns out that, provided $\frac{1}{2}<\rho\leq1$ and $\rho+\delta\geq1$,
there is an element of the quotient space $S_{\rho,\delta}^{m}/S_{\rho,\delta}^{m-\left(\rho-\delta\right)}$
called the \textbf{principal symbol} which is well-defined independently
of the charts. Any member of this equivalence class is a function
on the cotangent bundle $T^{*}M$, also called a principal symbol.
For technical reasons it is common and convenient to assume $\rho=1-\delta$
and $\rho>\frac{1}{2}$, and to denote $S_{\rho}^{m}=S_{\rho,\delta}^{m}$.
We will follow from now on this convention. 

It is well-known that many results of semi-classical analysis on a
manifold are given in terms of principal symbols. On the other hand,
when we consider symbols with a non-constant order function $m\left(x,\xi\right)\in T^{*}M$,
we need to manipulate carefully the concept of principal symbol, since
there are two different notions depending on whether we consider our
symbol to have a variable order $m\left(x,\xi\right)$ or a constant
order equal to $\limsup m$. 

To avoid possibly confusing considerations about principal symbols,
we will use a very convenient quantization scheme, developed in \cite{pflaum_2},
which works on Riemannian manifolds $\left(X,g\right)$ and provides
a notion of \textbf{total symbol}. It is defined as follows. First,
we fix a cut-off function $\chi\in C^{\infty}\left(TM,\left[0,1\right]\right)$
which equals $1$ on a neighborhood of the zero section $0_{TM}$
and is supported in a neighborhood $W\subset TM$ of $0_{TM}$ in
which the exponential map defines a diffeomorphism onto an open neighborhood
of the diagonal in $M\times M$. Then, for any $u\in C^{\infty}\left(M\right)$
we define $u_{\chi}\in C^{\infty}\left(TM\right)$ its semi-classical
lift on $TM$ by \[
u_{\chi}\left(x,v\right)=\left\{ \begin{array}{ll}
\chi\left(v\right)u\left(\exp_{x}v\right) & \mbox{for }\left(x,v\right)\in W\\
0 & \mbox{else.}\end{array}\right..\]
Finally, for any symbol $p\in C^{\infty}\left(T^{*}M\right)$ one
defines the operator $\hat{P}$ by \begin{equation}
u\rightarrow\hat{P}\left(u\right)\left(x\right)=\int_{T_{x}^{*}X}p\left(x,\xi\right)\widetilde{u_{\chi}}\left(x,\xi\right)d\xi\label{eq_pdo_on_manifold}\end{equation}
 where $\widetilde{f}$ denotes the Fourier transform of a function
$f\left(x,v\right)\in C^{\infty}\left(TM\right)$ with respect to
the $v$ variable, i.e.,\[
\widetilde{f}\left(x,\xi\right)=\frac{1}{\left(2\pi\right)^{n}}\int_{T_{x}X}e^{-i\left\langle \xi\mid v\right\rangle }f\left(x,v\right)dv.\]

It is shown in \cite{pflaum_2} that this construction gives rise
to a notion of total symbol (called there \textbf{normal symbol})
well-defined independently of the cut-off function $\chi$ up to an
element of $S_{\rho}^{-\infty}$. The classes of symbols are defined
in the usual way : Fix $m\in\mathbb{R}$ and $\rho>\frac{1}{2}$.
Then, a function $p\in C^{\infty}\left(T^{*}M\right)$ belongs to
the class $S_{\rho}^{m}$ if in any trivialization $\left(x,\xi\right):\left.T^{*}M\right|_{U}\rightarrow\mathbb{R}^{2n}$,
for any compact $K\subset U$ and any multi-indices $\alpha,\beta\in\mathbb{N}^{n}$,
there is a constant $C_{K,\alpha,\beta}$ such that \[
\left|\partial_{\xi}^{\alpha}\partial_{x}^{\beta}p\left(x,\xi\right)\right|\leq C_{K,\alpha,\beta}\left\langle \xi\right\rangle ^{m-\rho\left|\alpha\right|+\left(1-\rho\right)\left|\beta\right|}\]
on $\left.T^{*}M\right|_{U}$. Now the function $\left\langle \xi\right\rangle =\sqrt{1+\left|\xi\right|^{2}}$
is defined in term of the norm $\left|\xi\right|^{2}=g_{x}\left(\xi,\xi\right)$
with the scalar product on $T^{*}M$ denoted by the same letter $g$.

\subsection{Symbols with variable orders }

As usual, \emph{la constante positive} $C$\emph{, qui est le fidèle
compagnon de l'analyste, pourra varier d'une formule à l'autre}%
\footnote{A. Unterberger \cite{unterberger}%
}.

\subsubsection{Definition and main basic properties of $S_{\rho}^{m\left(x,\xi\right)}$}

As explained before, we want to develop a symbolic calculus for PDO's
whose symbol has an order $m$ which depends on the point $\left(x,\xi\right)\in T^{*}M$.
We first need to explain which functions are acceptable as order functions.

\vspace{0.cm}\begin{center}{\color{red}\fbox{\color{black}\parbox{16cm}{
\begin{Def}

\label{def_variable_order_function}An \textbf{order function} $m\left(x,\xi\right)$
is an element of $S_{r}^{0}$, for some $\frac{1}{2}<r<1$ which is
also bounded at infinity in $\xi$, i.e., \[
\sup_{x,\xi\in T^{*}M}\left|m\left(x,\xi\right)\right|<\infty.\]

\end{Def}
}}}\end{center}\vspace{0.cm}

We now define the class of symbols of variable order exactly in the
same way as symbols with constant order. We will use the slightly
abusive notation $S_{\rho}^{m\left(x,\xi\right)}$ to emphasize the
fact that $m\left(x,\xi\right)$ is not constant. 

\vspace{0.cm}\begin{center}{\color{red}\fbox{\color{black}\parbox{16cm}{
\begin{Def}

\label{def_symbol_class_variable_order}Let $m\left(x,\xi\right)\in S_{r}^{0}$
be an order function and $\frac{1}{2}<\rho<1$. A function $p\in C^{\infty}\left(T^{*}M\right)$
belongs to the class $S_{\rho}^{m\left(x,\xi\right)}$ if in any trivialization
$\left(x,\xi\right):\left.T^{*}M\right|_{U}\rightarrow\mathbb{R}^{2n}$,
for any compact $K\subset U$ and any multi-indices $\alpha,\beta\in\mathbb{N}^{n}$,
there is a constant $C_{K,\alpha,\beta}$ such that \[
\left|\partial_{\xi}^{\alpha}\partial_{x}^{\beta}p\left(x,\xi\right)\right|\leq C_{K,\alpha,\beta}\left\langle \xi\right\rangle ^{m\left(x,\xi\right)-\rho\left|\alpha\right|+\left(1-\rho\right)\left|\beta\right|}\]
for any $\left(x,\xi\right)\in\left.T^{*}M\right|_{U}$.

\end{Def}
}}}\end{center}\vspace{0.cm}

As with the usual calculus, for any symbol $p\in S_{\rho}^{m\left(x,\xi\right)}$
and any multi-indices $\alpha,\beta$ we have \[
\partial_{\xi}^{\alpha}\partial_{x}^{\beta}p\in S_{\rho}^{m\left(x,\xi\right)-\rho\left|\alpha\right|+\left(1-\rho\right)\left|\beta\right|}.\]
This means that $\partial_{\xi}^{\alpha}\partial_{x}^{\beta}p$ has
an order function given by $m\left(x,\xi\right)-\rho\left|\alpha\right|+\left(1-\rho\right)\left|\beta\right|$.
Similarly, for any two symbols $p\in S_{\rho}^{m\left(x,\xi\right)}$
and $q\in S_{\rho'}^{m'\left(x,\xi\right)}$, the point-wise product
$p\left(x,\xi\right)q\left(x,\xi\right)$ belongs to $S_{\rho''}^{m\left(x,\xi\right)+m'\left(x,\xi\right)}$
with $\rho''=\min\left(\rho,\rho'\right)$. 

Another simple but important property of variable order symbols, is
that they belong in fact to some Hörmander class. This follows from
the following fact. 

\vspace{0.cm}\begin{center}{\color{blue}\fbox{\color{black}\parbox{16cm}{
\begin{lem}

\label{lem_symb_variable_equal_symb_hormander}For any two non-constant
order $m\left(x,\xi\right)$ and $m'\left(x,\xi\right)$ satisfying
$m\left(x,\xi\right)\leq m'\left(x,\xi\right)$ on $T^{*}M$ and any
$\rho\geq\rho'$, the following holds \[
S_{\rho}^{m\left(x,\xi\right)}\subset S_{\rho'}^{m'\left(x,\xi\right)}.\]
In particular, one has

\[
S_{\rho}^{m\left(x,\xi\right)}\subset S_{\rho'}^{\sup m}\mbox{ and }S_{\rho}^{m\left(x,\xi\right)}\subset S_{\rho'}^{\varepsilon+\lim\sup m}\mbox{ for any }\varepsilon>0.\]

\end{lem}
}}}\end{center}\vspace{0.cm}

\begin{proof}
\small 

Let $p\in S_{\rho}^{m\left(x,\xi\right)}$ be a symbol. For any $\alpha,\beta\in\mathbb{N}^{n}$,
any compact $K\subset M$, the first inclusion comes simply from\[
\left|\partial_{\xi}^{\alpha}\partial_{x}^{\beta}p\left(x,\xi\right)\right|\leq C\left\langle \xi\right\rangle ^{m\left(x,\xi\right)-\rho\left|\alpha\right|+\left(1-\rho\right)\left|\beta\right|}\leq C\left\langle \xi\right\rangle ^{m'\left(x,\xi\right)-\rho'\left|\alpha\right|+\left(1-\rho'\right)\left|\beta\right|}.\]
On the other hand, for large enough $\left\langle \xi\right\rangle $,
$m$ is bounded by $\limsup m+\varepsilon$ for any $\varepsilon>0$.
This provides the second estimate.

\end{proof}
\normalsize

This property implies in particular that we can use the same class
of residual symbols as for Hörmander symbols, namely \[
S^{-\infty}:=\bigcap_{m<0}S_{\rho}^{m}\]
 which is independent of $\rho$. For convenience, we also introduce
the class of symbols of any order \[
S_{\rho}^{\infty}:=\bigcup_{m>0}S_{\rho}^{m}.\]

\subsubsection{Canonical examples }

We show now that the natural candidate $\left\langle \xi\right\rangle ^{m\left(x,\xi\right)}$
is indeed a suitable symbol. 

\vspace{0.cm}\begin{center}{\color{blue}\fbox{\color{black}\parbox{16cm}{
\begin{lem}

\label{lem_ksi_puissance_m_is_good_symbol}Let $m\in S_{\rho}^{0}$
be an order function. The smooth function $p\left(x,\xi\right)=\left\langle \xi\right\rangle ^{m\left(x,\xi\right)}$
belongs to $S_{\rho-\varepsilon}^{m\left(x,\xi\right)}$ for any $\varepsilon>0$.

\end{lem}
}}}\end{center}\vspace{0.cm}

\begin{proof}
\small 

We will prove that for any $\alpha,\beta$, we have \begin{equation}
\partial_{\xi}^{\alpha}\partial_{x}^{\beta}p\left(x,\xi\right)=q\left(x,\xi\right)\left\langle \xi\right\rangle ^{m\left(x,\xi\right)}\mbox{ with }q\in S_{\rho}^{\left(-\rho+\varepsilon\right)\left|\alpha\right|+\left(1-\rho+\varepsilon\right)\left|\beta\right|}\label{eq_good_symbol_1}\end{equation}
 for any $\varepsilon>0$. First of all, this is true for first order
derivatives. Indeed, for $\left|\beta\right|=1$ we compute the derivative
: \[
\partial_{x}^{\beta}p\left(x,\xi\right)=\left(\partial_{x}^{\beta}\left(\ln\left(\left\langle \xi\right\rangle \right)m\left(x,\xi\right)\right)\right)\left\langle \xi\right\rangle ^{m\left(x,\xi\right)}=:q\left(x,\xi\right)\left\langle \xi\right\rangle ^{m\left(x,\xi\right)}.\]
The appearance of logarithmic terms is actually the worse that can
happen when differentiating a symbol with variable order, but it is
easily controlled. First, $\ln\left(\left\langle \xi\right\rangle \right)$
is bounded by $\left\langle \xi\right\rangle ^{\varepsilon}$ for
$\varepsilon>0$ arbitrarily small. Moreover, the logarithm disappears
as soon as we take at least one derivative in $x$ or in $\xi$. Namely,
whenever $\left(\alpha,\beta\right)\neq\left(0,0\right)$ one has
$\partial_{\xi}^{\alpha}\partial_{x}^{\beta}\left(\ln\left(\left\langle \xi\right\rangle \right)\right)\in S_{1}^{-\left|\alpha\right|}$.
This shows that $\ln\left(\left\langle \xi\right\rangle \right)\in S_{1}^{\varepsilon}$
for any $\varepsilon>0$. This means that $\ln\left(\left\langle \xi\right\rangle \right)m\left(x,\xi\right)\in S_{1}^{\varepsilon}.S_{\rho}^{0}=S_{\rho}^{\varepsilon}$.
Then the derivative yields to $q\left(x,\xi\right)\in S_{\rho}^{\varepsilon+1-\rho}$.
Similarly, for $\left|\alpha\right|=1$ one shows that $\partial_{\xi}^{\alpha}p\left(x,\xi\right)=q\left(x,\xi\right)\left\langle \xi\right\rangle ^{m\left(x,\xi\right)}$
with $q\in S_{\rho}^{\varepsilon-\rho}$ for any $\varepsilon>0$.
Let us prove by iteration that Equation (\ref{eq_good_symbol_1})
holds in general. Suppose it holds for all $\alpha,\beta$ satisfying
$\left|\alpha+\beta\right|\leq N$ for some $N\in\mathbb{N}$. Then,
any $\left(\alpha',\beta'\right)$ with $\left|\alpha+\beta\right|=N+1$
has the form $\left(\alpha+a,\beta+b\right)$ with $\left|a+b\right|=1$,
i.e., either $\left(\left|a\right|,\left|b\right|\right)=\left(1,0\right)$
or $\left(\left|a\right|,\left|b\right|\right)=\left(0,1\right)$.
In the first case, we want to compute \[
\partial_{\xi}^{a}\partial_{\xi}^{\alpha}\partial_{x}^{\beta}p\left(x,\xi\right)=\left(\partial_{\xi}^{a}q\left(x,\xi\right)\right)\left\langle \xi\right\rangle ^{m\left(x,\xi\right)}+q\left(x,\xi\right)\partial_{\xi}^{a}\left(\left\langle \xi\right\rangle ^{m\left(x,\xi\right)}\right).\]
By assumption, the first term is in $S_{\rho}^{\left(-\rho+\varepsilon\right)\left|\alpha\right|+\left(1-\rho+\varepsilon\right)\left|\beta\right|-\rho}.\left\langle \xi\right\rangle ^{m\left(x,\xi\right)}$
and the second one is in \[
S_{\rho}^{\left(-\rho+\varepsilon\right)\left|\alpha\right|+\left(1-\rho+\varepsilon\right)\left|\beta\right|}.S_{\rho}^{-\rho+\varepsilon}.\left\langle \xi\right\rangle ^{m\left(x,\xi\right)}.\]
Together, this provides \[
\left\langle \xi\right\rangle ^{-m\left(x,\xi\right)}\partial_{\xi}^{a}\partial_{\xi}^{\alpha}\partial_{x}^{\beta}p\left(x,\xi\right)\in S_{\rho}^{\left(-\rho+\varepsilon\right)\left|\alpha+a\right|+\left(1-\rho+\varepsilon\right)\left|\beta\right|}.\]
Similarly, we would find \[
\left\langle \xi\right\rangle ^{m\left(x,\xi\right)}\partial_{x}^{b}\partial_{\xi}^{\alpha}\partial_{x}^{\beta}p\left(x,\xi\right)\in S_{\rho}^{\left(-\rho+\varepsilon\right)\left|\alpha\right|+\left(1-\rho+\varepsilon\right)\left|\beta+b\right|}\]
 for $\left|b\right|=1$. This proves Formula (\ref{eq_good_symbol_1})
for any $\alpha,\beta$ satisfying $\left|\alpha+\beta\right|\leq N+1$
and the formula for all $\alpha,\beta$ is proved by induction. Then,
we deduce that 

\[
\left|\partial_{\xi}^{\alpha}\partial_{x}^{\beta}p\left(x,\xi\right)\right|\leq\left\langle \xi\right\rangle ^{m\left(x,\xi\right)-\left(\rho-\varepsilon\right)\left|\alpha\right|+\left(1-\rho+\varepsilon\right)\left|\beta\right|}\]
 and thus $p\left(x,\xi\right)\in S_{\rho-\varepsilon}^{m\left(x,\xi\right)}$
for any $\varepsilon>0$. 

\end{proof}
\normalsize

\subsubsection{Action of diffeomorphisms}

For any diffeomorphism $\phi:M\rightarrow M$, we denote by $\phi_{*}:T^{*}M\rightarrow T^{*}M$
the lift on the cotangent bundle defined by $\phi_{*}\left(x,\xi\right)=\left(\phi\left(x\right),\left(\left(D_{x}\phi\right)^{-1}\right)^{t}\right)$. 

\vspace{0.cm}\begin{center}{\color{blue}\fbox{\color{black}\parbox{16cm}{
\begin{lem}

\label{lem_symbole_variable_invariance_diffeo}Let $p\in S_{\rho}^{m\left(x,\xi\right)}$
be a symbol with non-constant order $m\left(x,\xi\right)$ and $\phi:M\rightarrow M$
a diffeomorphism. Then, the composition $p\circ\phi_{*}$ belongs
to $S_{\rho}^{m\circ\phi_{*}}$.

\end{lem}
}}}\end{center}\vspace{0.cm}

\begin{proof}
\small 

Set $a=p\circ\phi_{*}$. For simplicity, we will write $\xi^{a}\circ\phi_{*}$
for the $\xi^{a}$ component of $\phi_{*}\left(x,\xi\right)$. We
will prove by iteration that for any order function $m\left(x,\xi\right)$,
any symbol $p\in S_{\rho}^{m}$ and any $\alpha,\beta\in\mathbb{N}^{d}$,
one has \begin{equation}
\partial_{\xi}^{\alpha}\partial_{x}^{\beta}\left(p\circ\phi_{*}\right)=q_{\alpha,\beta}\circ\phi_{*}\mbox{ where }q_{\alpha,\beta}\in S_{\rho}^{m-\rho\left|\alpha\right|+\left(1-\rho\right)\left|\beta\right|},\label{eq_symb_invar_diffeo_1}\end{equation}
where we write $m=m\left(x,\xi\right)$ for shortness. First of all,
this is trivially true for $\alpha=\beta=0$. Then, we suppose it
is true for all $\alpha,\beta\in\mathbb{N}^{d}$, with $\left|\alpha+\beta\right|\leq N$.
If we compute the derivative $\partial_{\xi}^{\alpha+a}\partial_{x}^{\beta}$
of $p\circ\phi_{*}$ with $\left|a\right|=1$, we obtain simply \begin{eqnarray*}
\partial_{\xi}^{\alpha+a}\partial_{x}^{\beta}\left(p\circ\phi_{*}\right) & = & \partial_{\xi}^{a}\left(q_{\alpha,\beta}\circ\phi_{*}\right)\\
 & = & \sum_{\left|a'\right|=1}\partial_{\xi}^{a'}\left(q_{\alpha,\beta}\right)\circ\phi_{*}.\partial_{\xi}^{a}\left(\xi^{a'}\circ\phi_{*}\right).\end{eqnarray*}
The terms $\partial_{\xi}^{a'}\left(q_{\alpha,\beta}\right)$ live
in $S_{\rho}^{m-\rho\left|\alpha+a\right|+\left(1-\rho\right)\left|\beta\right|}$
by assumption whereas the second term in the product belongs to $S_{1}^{0}$.
This means that $\partial_{\xi}^{\alpha+a}\partial_{x}^{\beta}\left(p\circ\phi_{*}\right)=q_{\alpha+a,\beta}\circ\phi_{*}$
where the symbol \[
q_{\alpha+a,\beta}=\sum_{\left|a'\right|=1}\partial_{\xi}^{a'}\left(q_{\alpha,\beta}\right).\partial_{\xi}^{a}\left(\xi^{a'}\circ\phi_{*}\right)\circ\phi_{*}^{-1}\]
 belongs to $S_{\rho}^{m-\rho\left|\alpha+a\right|+\left(1-\rho\right)\left|\beta\right|}$.
We consider now the derivative $\partial_{\xi}^{\alpha}\partial_{x}^{\beta+b}$
with $\left|b\right|=1$. The computation is slightly more complicated,
since there is an $x$-dependence in the $\xi$ component of $\phi_{*}\left(x,\xi\right)$.
Namely, \begin{eqnarray*}
\partial_{\xi}^{\alpha}\partial_{x}^{\beta+b}\left(p\circ\phi_{*}\right) & = & \sum_{\left|b'\right|=1}\partial_{x}^{b'}\left(q_{\alpha,\beta}\right)\circ\phi_{*}.\partial_{x}^{b}\left(\phi^{b'}\left(x\right)\right)\\
 &  & +\sum_{\left|a'\right|=1}\partial_{\xi}^{a'}\left(q_{\alpha,\beta}\right)\circ\phi_{*}.\partial_{x}^{b}\left(\xi^{a'}\circ\phi_{*}\right).\end{eqnarray*}
In the first sum, the terms $\partial_{x}^{b'}\left(q_{\alpha,\beta}\right)$
belong to $S_{\rho}^{m-\rho\left|\alpha\right|+\left(1-\rho\right)\left|\beta+b\right|}$
and $\partial_{x}^{b}\left(\phi^{b'}\left(x\right)\right)$ is in
$S_{1}^{0}$. On the other hand, in the second sum we have $\partial_{\xi}^{a'}\left(q_{\alpha,\beta}\right)\in S_{\rho}^{m-\rho\left|\alpha+a\right|+\left(1-\rho\right)\left|\beta\right|}$
but the second one in the product is in $S_{1}^{1}$. This means that
the second sum brings a power $\left(1-\rho\right)$ of $\xi$. All
together, we obtain $\partial_{\xi}^{\alpha}\partial_{x}^{\beta+b}\left(p\circ\phi_{*}\right)=q_{\alpha,\beta+b}\circ\phi_{*}$
with $q_{\alpha,\beta+b}\in S_{\rho}^{m-\rho\left|\alpha+a\right|+\left(1-\rho\right)\left|\beta\right|+1-\rho}$.
This proves therefore by induction Formula (\ref{eq_symb_invar_diffeo_1})
for all $\alpha,\beta\in\mathbb{N}^{d}$. 

Now, the lemma follows easily form this formula. Indeed, for any $\alpha,\beta\in\mathbb{N}^{d}$
one has \[
\left|\partial_{\xi}^{\alpha}\partial_{x}^{\beta}\left(p\circ\phi_{*}\right)\right|\leq C\left\langle \xi\circ\phi_{*}\right\rangle ^{m\circ\phi_{*}-\rho\left|\alpha\right|+\left(1-\rho\right)\left|\beta\right|}.\]
On the other hand, since both $D_{x}\phi$ and $\left(D_{x}\phi\right)^{-1}$
are uniformly bounded on $M$, it follows that there is a constant
$C>1$ such that $\frac{1}{C}\left\langle \xi\right\rangle \leq\left\langle \xi\circ\phi_{*}\right\rangle \leq C\left\langle \xi\right\rangle $.
This implies that \[
\left|\partial_{\xi}^{\alpha}\partial_{x}^{\beta}\left(p\circ\phi_{*}\right)\right|\leq C\left\langle \xi\right\rangle ^{m\circ\phi_{*}-\rho\left|\alpha\right|+\left(1-\rho\right)\left|\beta\right|}.\]

\end{proof}
\normalsize

\subsection{PDO with variable order}

Given an order function $m\left(x,\xi\right)$ and a symbol $p\left(x,\xi\right)$
in the class $S_{\rho}^{m\left(x,\xi\right)}$, Formula (\ref{eq_pdo_on_manifold})
provides an operator from $C_{0}^{\infty}\left(M\right)$ to $C^{\infty}\left(M\right)$,
which is actually continuous. By duality, it is also continuous from
$\mathcal{E}'$ to $\mathcal{D}'$. The class $\Psi_{\rho}^{m\left(x,\xi\right)}$
of PDO's is then the set of operators of the form (\ref{eq_pdo_on_manifold})
modulo a \textbf{smoothing} operator, i.e. an operator which sends
$\mathcal{E}'$ into $C^{\infty}\left(M\right)$ continuously. We
denote by $\Psi^{-\infty}=\bigcap_{n>0}\Psi_{\rho}^{-n}$ the class
of smoothing operators and also the class $\Psi_{\rho}^{\infty}=\bigcup_{n>0}\Psi_{\rho}^{n}$
of all PDOs of type $\rho$ which contains $\Psi_{\rho}^{m\left(x,\xi\right)}$
for any variable order $m\left(x,\xi\right)$. Notice that given an
operator $\hat{p}\in\Psi_{\rho}^{\infty}$, its symbol is well-defined
only up to an element in $S_{\rho}^{-\infty}$. 

We now review the most important properties of PDO's. The proofs for
non-constant order symbols follow in most cases the line of the proofs
for usual symbols (see for example \cite{grigis_sjostrand,taylor_tome2})
and are omitted for shortness of this paper. In all the sequel, the
parameter $\rho$ is always supposed to satisfy $\rho>\frac{1}{2}$.
As well, in order to avoid any discussion about properly supported
operators, we assume from now on $M$ to be compact, since it will
be the case for the application of these tools to Ruelle-Pollicott
resonances.

\subsubsection{Asymptotic expansions }

Semi-classical analysis is naturally an asymptotic theory. In order
to prove the basic theorems about composition, Egorov, ellipticity
or functional calculus, one needs to give a sense to formal series
like $\sum p_{j}$, where $\left\{ p_{j}\right\} _{j\in\mathbb{N}}$
is a sequence of symbols with decreasing orders. Such a series is
most of the time divergent, but it is possible to find a symbol $p$
which is asymptotically equivalent to the series. This is an adaptation
of an old result by Borel.

\vspace{0.cm}\begin{center}{\color{blue}\fbox{\color{black}\parbox{16cm}{
\begin{theo}

\label{thm:borel_resummation}Let $p_{j}\in S_{\rho}^{m_{j}\left(x,\xi\right)}$
be a sequence of symbols with variable order $m_{j}\in S_{r}^{1}$,
with $\rho<r\leq1$, satisfying $m_{j}\downarrow-\infty$, in the
sense that, for all $j\in\mathbb{N}$ \[
\sup_{x,\xi}m_{j}\left(x,\xi\right)\rightarrow-\infty\mbox{ and }m_{j+1}\left(x,\xi\right)\leq m_{j}\left(x,\xi\right).\]
 Then, there exists a symbol $p\in S_{\rho}^{m_{0}\left(x,\xi\right)}$
such that for all $N\geq0$ \[
p-\sum_{j=0}^{N-1}p_{j}\in S_{\rho}^{m_{N}\left(x,\xi\right)}.\]
The symbol $p$ is unique modulo a residual symbol, i.e., an element
of $S^{-\infty}$.

\end{theo}
}}}\end{center}\vspace{0.cm}

The proof of this theorem is a straightforward adaptation of the proof
for usual symbols, which can be found for example in \cite[II, 3]{martinez-01}
or \cite{grigis_sjostrand}. This fact implies automatically the corresponding
result for asymptotic sums of PDO's. Namely, if $\hat{p}_{j}\in\Psi_{\rho}^{m_{j}\left(x,\xi\right)}$
is sequence of PDO with decreasing orders, then there exist an operator
$\hat{p}\in\Psi_{\rho}^{m_{0}\left(x,\xi\right)}$ which satisfies
\[
\hat{p}-\sum_{j=0}^{N-1}\hat{p}_{j}\in\Psi_{\rho}^{m_{N}\left(x,\xi\right)}\]
for all $N\geq0$.

\subsubsection{Adjoint and composition}

\vspace{0.cm}\begin{center}{\color{blue}\fbox{\color{black}\parbox{16cm}{
\begin{theo}

\label{thm_adjoint}Let $\hat{p}\in\Psi_{\rho}^{m\left(x,\xi\right)}$
be a PDO with non-constant order symbol $p\in S_{\rho}^{m\left(x,\xi\right)}$.
Then the adjoint $\hat{p}^{*}$ is itself a PDO in $\Psi_{\rho}^{m\left(x,\xi\right)}$
and its symbol $p^{*}\in S_{\rho}^{m\left(x,\xi\right)}$ satisfies
\[
p^{*}\left(x,\xi\right)-\overline{p\left(x,\xi\right)}\in S_{\rho}^{m\left(x,\xi\right)-\left(2\rho-1\right)},\]
where $\overline{{\normalcolor \phantom{z}}}$ denotes the complex
conjugate. 

\end{theo}
}}}\end{center}\vspace{0.cm}

Here, the adjoint means the formal $L^{2}$-adjoint defined on the
same domain $C^{\infty}\left(M\right)$. 

\vspace{0.cm}\begin{center}{\color{blue}\fbox{\color{black}\parbox{16cm}{
\begin{theo}

\label{thm_composition}Let $\hat{p}\in\Psi_{\rho}^{m\left(x,\xi\right)}$
and $\hat{q}\in\Psi_{\rho}^{m'\left(x,\xi\right)}$ be two PDO's with
non-constant order $m\left(x,\xi\right)$ and $m'\left(x,\xi\right)$.
Then the product $\hat{a}:=\hat{p}\hat{q}$ is a PDO in $\Psi_{\rho}^{m\left(x,\xi\right)+m'\left(x,\xi\right)}$
and its symbol $a\left(x,\xi\right)$ satisfies \[
a\left(x,\xi\right)-p\left(x,\xi\right)q\left(x,\xi\right)\in S_{\rho}^{m\left(x,\xi\right)+m'\left(x,\xi\right)-\left(2\rho-1\right)}.\]

\end{theo}
}}}\end{center}\vspace{0.cm}

\subsubsection{Egorov's theorem }

Egorov's Theorem describes how PDO's transform under conjugation with
a Fourier Integral Operator. We will nevertheless avoid talking about
general FIO's and restrict ourselves to the simplest case, namely
the composition by a diffeomorphism on $M$, which is sufficient for
our purposes. See \cite[p.24]{taylor_tome2}

\vspace{0.cm}\begin{center}{\color{blue}\fbox{\color{black}\parbox{16cm}{
\begin{theo}

\label{thm_egorov}Let $\hat{p}\in\Psi_{\rho}^{m\left(x,\xi\right)}$
be a PDO with non-constant order $m\left(x,\xi\right)$ and $f:M\rightarrow M$
a diffeomorphism. Denote by $\hat{F}$ the pull-back operator $\hat{F}\left(u\right)=u\circ f$
and by $F:T^{*}M\rightarrow T^{*}M$ the lift of $f^{-1}$ to the
cotangent bundle defined by $F\left(x,\xi\right)=\left(f^{-1}\left(x\right),\left(D_{x}f\right)^{t}\right)$.
Then, the conjugation $\hat{a}:=\hat{F}^{-1}\hat{p}\hat{F}$ belongs
to $\Psi_{\rho}^{m\circ F\left(x,\xi\right)}$ and its symbol $a\left(x,\xi\right)$
satisfies \[
a\left(x,\xi\right)-p\circ F\left(x,\xi\right)\in S_{\rho}^{m\circ F\left(x,\xi\right)-\left(2\rho-1\right)}.\]

\end{theo}
}}}\end{center}\vspace{0.cm}

\subsubsection{Sobolev continuity}

It is a well-known fact that on a compact manifold, a PDO of constant
order $m$ extends to a continuous operator $H^{s}\rightarrow H^{s-m}$
for all $s$. Thanks to Lemma \ref{lem_symb_variable_equal_symb_hormander},
a PDO with variable order $m\left(x,\xi\right)$ extends to a continuous
operators $H^{s}\rightarrow H^{s-m^{+}}$ with $m^{+}=\limsup m\left(x,\xi\right)$.
On the other hand the embeddings $H^{s}\hookrightarrow H^{s'}$ for
$s'<s$ are compact. In particular, smoothing operators are compact
in any Sobolev space.

\subsection{Non-isotropic ellipticity }

\subsubsection{Variable order ellipticity }

We know from the standard theory of PDO's that an operator $\hat{p}\in\Psi_{\rho}^{m}$
is invertible modulo $\Psi^{-\infty}$ with an {}``inverse'' in
$\Psi_{\rho}^{-m}$ as soon as its symbol $p$ satisfies an \textbf{ellipticity}
condition. We now show that the classical definition of ellipticity
extends in a natural way to symbols with variable order $m\left(x,\xi\right)$.
This leads to a more general notion of ellipticity which is proved
to be equivalent to the existence of a parametrix $\Psi_{\rho}^{-m\left(x,\xi\right)}$.

\vspace{0.cm}\begin{center}{\color{red}\fbox{\color{black}\parbox{16cm}{
\begin{Def}

\label{def_elliptic}A symbol a non-constant order $p\in S_{\rho}^{m\left(x,\xi\right)}$
is called \textbf{elliptic} if there is a $C>0$ such that $\left|p\left(x,\xi\right)\right|\geq\frac{1}{C}\left\langle \xi\right\rangle ^{m\left(x,\xi\right)}$
whenever $\left\langle \xi\right\rangle \geq C$. An operator $\hat{p}$
is elliptic if its symbol is elliptic. 

\end{Def}
}}}\end{center}\vspace{0.cm}

One can easily check that the following statement is equivalent to
ellipticity.

\vspace{0.cm}\begin{center}{\color{blue}\fbox{\color{black}\parbox{16cm}{
\begin{lem}

\label{lem_elliptic_equivalent}A symbol $p\in S_{\rho}^{m\left(x,\xi\right)}$
is elliptic if and only if there exists a symbol $q\in S_{\rho}^{-m\left(x,\xi\right)}$
such that \[
p\left(x,\xi\right)q\left(x,\xi\right)-1\in S_{\rho}^{-\infty}.\]

\end{lem}
}}}\end{center}\vspace{0.cm}

\begin{example}

\label{exa_elliptic}For any order function $m\left(x,\xi\right)$
the symbol $p\left(x,\xi\right)=\left\langle \xi\right\rangle ^{m\left(x,\xi\right)}$
is elliptic and one can choose $q=\left\langle \xi\right\rangle ^{-m\left(x,\xi\right)}$
on the whole of $T^{*}M$. 

\end{example}

For usual symbols, it is well-known that ellipticity is a phenomenon
of the principal symbol. This is also true for variable order symbols,
in the following sense.

\vspace{0.cm}\begin{center}{\color{blue}\fbox{\color{black}\parbox{16cm}{
\begin{lem}

\label{lem_elliptic_principal_symbol}Let $p\in S_{\rho}^{m\left(x,\xi\right)}$
be an elliptic symbol. Then any other symbol $q\in S_{\rho}^{m\left(x,\xi\right)}$
satisfying $p-q\in S_{\rho}^{m\left(x,\xi\right)-\varepsilon}$ for
some $\varepsilon>0$ is elliptic as-well.

\end{lem}
}}}\end{center}\vspace{0.cm}

\begin{proof}
\small 

Suppose $p=q+s$ with $s\in S_{\rho}^{m\left(x,\xi\right)-\varepsilon}$.
Ellipticity of $p$ means that for large enough $\left\langle \xi\right\rangle $
one has $\left|p\left(x,\xi\right)\right|\geq\frac{1}{C}\left\langle \xi\right\rangle ^{m\left(x,\xi\right)}$.
On the other hand, for large $\left\langle \xi\right\rangle $ one
has also $\left|s\left(x,\xi\right)\right|\leq C\left\langle \xi\right\rangle ^{m\left(x,\xi\right)-\varepsilon}$.
Therefore \[
\left|q\left(x,\xi\right)\right|\geq C\left\langle \xi\right\rangle ^{m\left(x,\xi\right)}\left(1-C'\left\langle \xi\right\rangle ^{-\varepsilon}\right)\geq C''\left\langle \xi\right\rangle ^{m\left(x,\xi\right)}\]
 for large $\left\langle \xi\right\rangle $, hence $q$ is elliptic.

\end{proof}
\normalsize

Notice that this notion of ellipticity is more general than the usual
one. Indeed, the symbol $\left\langle \xi\right\rangle ^{m\left(x,\xi\right)}$
with an order which takes its values, say between $-1$ and $+1$
is not elliptic in the usual sense. Indeed, the symbol $\left\langle \xi\right\rangle ^{m\left(x,\xi\right)}$
with an order taking its values, say between -1 and +1 is not elliptic
in the usual sense, when we view it as a symbol with constant order
$\sup m$.

\subsubsection{Parametrix and invertibility}

The main point in considering this notion of non-isotropic ellipticity
is of course that it is equivalent to the existence of a parametrix,
as explained in Theorem \ref{thm_elliptic-parametrix} below.

\vspace{0.cm}\begin{center}{\color{red}\fbox{\color{black}\parbox{16cm}{
\begin{Def}

\label{def_parametrix}Let $\hat{p}\in\Psi_{\rho}^{m\left(x,\xi\right)}$
be any PDO. A \textbf{parametrix} of $\hat{p}$ is a PDO $\hat{q}\in\Psi_{\rho}^{-m\left(x,\xi\right)}$
such that \[
\hat{p}\hat{q}-\mathbb{I}\in\Psi^{-\infty}\mbox{ and }\hat{q}\hat{p}-\mathbb{I}\in\Psi^{-\infty}.\]

\end{Def}
}}}\end{center}\vspace{0.cm}

\vspace{0.cm}\begin{center}{\color{blue}\fbox{\color{black}\parbox{16cm}{
\begin{theo}

\label{thm_elliptic-parametrix}An operator $\hat{p}\in\Psi_{\rho}^{m\left(x,\xi\right)}$
admits a parametrix if and only if its symbol $p$ is elliptic.

\end{theo}
}}}\end{center}\vspace{0.cm}

For this reason, we will say equally that the symbol $p$ or the operator
$\hat{p}$ is elliptic. 

The construction is standard. We just check that it works as-well
in the variable order context. Assume $p\in S_{\rho}^{m\left(x,\xi\right)}$
is elliptic. Lemma \ref{lem_elliptic_equivalent} implies that there
is a $q_{0}\in S_{\rho}^{-m\left(x,\xi\right)}$ such that $\hat{p}\hat{q}_{0}=\mathbb{I}-\hat{r}$
where $\hat{r}\in\Psi_{\rho}^{-\left(2\rho-1\right)}$ because of
Theorem \ref{thm_composition}. This implies $\hat{r}^{j}\in\Psi_{\rho}^{-j\left(2\rho-1\right)}$
for all $j\in\mathbb{N}$ and it follows that $\hat{q}_{N}=\hat{q}_{0}\left(\mathbb{I}+\hat{r}+...+\hat{r}^{N}\right)$
satisfies \[
\hat{p}\hat{q}_{N-1}-\mathbb{I}\in\Psi_{\rho}^{-N\left(2\rho-1\right)}.\]
On the other hand, thanks to the re-summation Theorem \ref{thm:borel_resummation}
we can find a $\hat{q}_{R}\in\Psi_{\rho}^{-m\left(x,\xi\right)}$
satisfying $\hat{q_{R}}-\hat{q}_{N-1}\in\Psi_{\rho}^{-m\left(x,\xi\right)-N\left(2\rho-1\right)}$
for all $N\in\mathbb{N}$. Therefore we have\[
\hat{p}\hat{q}_{R}-\mathbb{I}\in\Psi_{\rho}^{-N\left(2\rho-1\right)}\]
 for all $N\in\mathbb{N}$, hence $\hat{q}_{R}$ is a right parametrix
for $\hat{p}$. Similarly, we can construct a left parametrix $\hat{q}_{L}\sim\left(\mathbb{I}+\hat{s}+\hat{s}^{2}+...\right)\hat{q}_{0}$
with $\hat{s}\in\Psi_{\rho}^{-\left(2\rho-1\right)}$ given by $\hat{p}\hat{q}_{0}=\mathbb{I}-\hat{s}$.
Finally, the fact that $\hat{q}_{L}-\hat{q}_{R}\in\Psi_{\rho}^{-\infty}$
comes from the observation that both $\hat{q}_{L}\hat{p}\hat{q}_{R}-\hat{q}_{R}$
and $\hat{q}_{L}\hat{p}\hat{q}_{R}-\hat{q}_{L}$ are smoothing. Therefore,
say $\hat{q}_{L}$ is a (both sided) parametrix for $\hat{p}$. 

\begin{proof}
\small 

\end{proof}
\normalsize

The existence of a parametrix has many interesting consequences, such
as those listed below. First of all, it is well-known that a standard
elliptic PDO (with constant order $m$) is Fredholm $H^{s}\rightarrow H^{s-m}$
for any $s$. For PDO's with variable order, a slightly weaker result
holds. 

\vspace{0.cm}\begin{center}{\color{blue}\fbox{\color{black}\parbox{16cm}{
\begin{lem}

\label{lem_elliptic_fredholm}Let $\hat{p}\in\Psi_{\rho}^{m\left(x,\xi\right)}$
be elliptic. Then the kernel of the operator $\hat{p}:H^{s}\rightarrow H^{s-m^{+}}$
with $m^{+}=\limsup m\left(x,\xi\right)$ is finite dimensional and
contained in $C^{\infty}\left(M\right)$. 

\end{lem}
}}}\end{center}\vspace{0.cm}

\begin{proof}
\small 

The key point in this proof is the fact that for any smoothing operator
$\hat{r}\in\Psi^{-\infty}$, the operator $\mathbb{I}+\hat{r}:H^{s}\rightarrow H^{s}$
is Fredholm for any $s$ and its kernel is contained in $C^{\infty}\left(M\right)$.
See for example \cite[ch. 7]{melrose_cours} for a proof for $s=0$
which extends straightforwardly to the case $s\neq0$. 

Now, the ellipticity $\hat{p}$ implies the existence of a left parametrix
$\hat{q}\in\Psi_{\rho}^{-m\left(x,\xi\right)}$, i.e.  $\hat{q}\hat{p}=\mathbb{I}+\hat{r}$
with $\hat{r}\in\Psi^{-\infty}$. This operator extends to $\hat{q}:H^{s-m^{+}}\rightarrow H^{s-m^{+}-m^{-}}$
where $m^{-}=\limsup\left(-m\left(x,\xi\right)\right)$. It follows
that $\ker\hat{p}$ is contained in the kernel of \[
\mathbb{I}+\hat{r}_{1}:H^{s}\rightarrow H^{s}\subset H^{s-m^{+}-m^{-}}\]
 which is finite dimensional and itself contained in $C^{\infty}\left(M\right)$. 

\end{proof}
\normalsize

This lemma has the consequence that we can make $\hat{p}$ invertible
by adding a smoothing operator, as shown in Lemma \ref{lem_elliptic_invertible}.
This is useful in practice, since one often needs to construct PDO's
whose symbol satisfies certain properties which are not modified by
adding a residual term. One needs first a preliminary result. 

\vspace{0.cm}\begin{center}{\color{blue}\fbox{\color{black}\parbox{16cm}{
\begin{lem}

\label{lem_elliptic_self_adjoint}Let $\hat{p}\in\Psi_{\rho}^{m\left(x,\xi\right)}$
be an elliptic and formally self-adjoint operator. Then $\hat{p}$
viewed as an unbounded operator on $L^{2}$ admits a self-adjoint
extension. 

\end{lem}
}}}\end{center}\vspace{0.cm}

\begin{proof}
\small 

The formal self-adjointness of $\hat{p}$ implies according to Theorem
\ref{thm_adjoint} that its symbol satisfies \[
p-Re\left(p\right)\in S_{\rho}^{m\left(x,\xi\right)-\left(2\rho-1\right)},\]
hence $\left|Re\left(p\left(x,\xi\right)\right)\right|\geq c.\left|p\left(x,\xi\right)\right|$,
with $c>0$, for large enough $\left|\xi\right|$. We can thus suppose
that $Re\left(p\left(x,\xi\right)\right)>0$ for large enough $\left|\xi\right|$
(if $Re\left(p\right)$ has the opposite sign, then the following
argument applies to $-\hat{p}$). We can therefore apply Lemma \ref{lem_square_root}
of the next section to show that \[
\hat{p}=\hat{b}^{*}\hat{b}-\hat{K}\]
 with $\hat{b}\in\Psi_{\rho}^{\frac{1}{2}m\left(x,\xi\right)}$ and
$\hat{K}\in\Psi^{-\infty}$. Since $\hat{K}$ is bounded in $L^{2}$,
it follows that $\hat{p}$ is bounded from below in $L^{2}$:\[
\left(\hat{p}u,u\right)\geq\left|\hat{b}u\right|^{2}-\left|\hat{K}\right|\left|u\right|^{2}.\]
This allows us to construct the Friedrichs extension of $\hat{p}$,
which is self-adjoint on a domain in $L^{2}$ (see e.g. \cite[p. 317]{yosida}).

\end{proof}
\normalsize

\vspace{0.cm}\begin{center}{\color{blue}\fbox{\color{black}\parbox{16cm}{
\begin{lem}

\label{lem_elliptic_invertible}Let $\hat{p}\in\Psi_{\rho}^{m\left(x,\xi\right)}$
be an elliptic and formally self-adjoint operator. Then there exists
a formally self-adjoint and smoothing operator $\hat{r}\in\Psi_{\rho}^{-\infty}$
such that $\hat{p}+\hat{r}$ is invertible $C^{\infty}\left(M\right)\rightarrow C^{\infty}\left(M\right)$
with inverse in $\Psi_{\rho}^{-m\left(x,\xi\right)}$.

\end{lem}
}}}\end{center}\vspace{0.cm}

\begin{proof}
\small 

This is also a standard construction. First, Lemma \ref{lem_elliptic_fredholm}
tells us that $\hat{p}:H^{m^{+}}\rightarrow L^{2}$ with $m^{+}=\limsup m\left(x,\xi\right)$
has a finite dimensional kernel contained in $C^{\infty}\left(M\right)$.
But this implies that viewed as an unbounded operator on $L^{2}$
with domain $H^{m^{+}}$, the operator $\hat{p}$ has also a finite
dimensional kernel contained in $C^{\infty}\left(M\right)$. On the
other hand, $\hat{p}$ has a self-adjoint extension on $L^{2}$ thanks
to Lemma \ref{lem_elliptic_self_adjoint}. Denote by \emph{$\mathcal{D}$}
the domain of this extension. This leads to the orthogonal decomposition
$L^{2}=\overline{\mbox{im}\left(\hat{p}\right)}\overset{\perp}{\oplus}\ker\left(\hat{p}\right)$
and the restriction $\hat{p}:\overline{\mbox{im}\left(\hat{p}\right)}\cap\mathcal{D}\rightarrow\mbox{im}\left(\hat{p}\right)$
is thus invertible. Therefore, the operator $\hat{P}$ given in matrix
form by \[
\hat{P}:=\left(\begin{array}{cc}
\left.\hat{p}\right|_{\overline{\mbox{im}\left(\hat{p}\right)}\cap\mathcal{D}} & 0\\
0 & \mathbb{I}\end{array}\right)\]
is invertible. We first remark that $\hat{P}$ is related to $\hat{p}$
by \[
\hat{P}=\hat{p}\left(1-\pi\right)+\pi\]
with $\pi:L^{2}\rightarrow\ker\hat{p}$ the orthogonal $L^{2}$-projection.
Since $\ker p$ is a finite dimensional subspace of $C^{\infty}\left(M\right)$,
the projection $\pi$ is a smoothing operator. It follows that $\hat{r}:=\hat{P}-\hat{p}=\left(1-\hat{p}\right)\pi$
is self-adjoint and smoothing. In particular $\hat{P}\in\Psi_{\rho}^{m\left(x,\xi\right)}$
is a PDO and defines an injective map $C^{\infty}\left(M\right)\rightarrow C^{\infty}\left(M\right)$.
On the other hand, the existence of a parametrix $\hat{Q}\in\Psi_{\rho}^{-m\left(x,\xi\right)}$
for $\hat{p}$, and thus for $\hat{P}$, implies that $\hat{P}$ is
also surjective $C^{\infty}\left(M\right)\rightarrow C^{\infty}\left(M\right)$.
Finally, denote by $\hat{P}^{-1}:C^{\infty}\left(M\right)\rightarrow C^{\infty}\left(M\right)$
the inverse of $\hat{P}$ which is continuous by the open mapping
Theorem. One has \[
\hat{Q}=\hat{Q}\hat{P}\hat{P}^{-1}=\hat{P}^{-1}+\hat{r}\hat{P}^{-1}\]
 with $\hat{r}$ smoothing. The last term is also smoothing since
$\hat{P}^{-1}$ is continuous and $\hat{r}$ smoothing. This implies
that $\hat{P}^{-1}$ is itself a PDO in $\Psi_{\rho}^{-m\left(x,\xi\right)}$.

\end{proof}
\normalsize

Collecting the results of this section, we obtain the following corollary. 

\vspace{0.cm}\begin{center}{\color{blue}\fbox{\color{black}\parbox{16cm}{
\begin{coro}

\label{cor_quantif_aa_inversible}For any real elliptic symbol $q\in S_{\rho}^{m\left(x,\xi\right)}$,
there is an operator $\hat{p}\in\Psi_{\rho}^{m\left(x,\xi\right)}$
satisfying $\hat{p}-\hat{q}\in\Psi_{\rho}^{m\left(x,\xi\right)-\left(2\rho-1\right)}$,
which is formally self-adjoint and invertible $C^{\infty}\left(M\right)\rightarrow C^{\infty}\left(M\right)$.

\end{coro}
}}}\end{center}\vspace{0.cm}

\begin{proof}
\small 

Let $\hat{q}\in\Psi_{\rho}^{m\left(x,\xi\right)}$ be the quantized
of the symbol $q$ and take the real part $\hat{a}:=\frac{1}{2}\left(\hat{q}+\hat{q}^{*}\right)$.
It is self-adjoint and according to Theorem \ref{thm_adjoint}, its
symbol satisfies $a\left(x,\xi\right)=q\left(x,\xi\right)$ mod $S_{\rho}^{m\left(x,\xi\right)-\left(2\rho-1\right)}$.
Then, Lemma \ref{lem_elliptic_principal_symbol} implies that the
ellipticity of $q$ is not destroyed by a modification of order $m\left(x,\xi\right)-\left(2\rho-1\right)$.
Therefore $a\in S_{\rho}^{m\left(x,\xi\right)}$ is elliptic as-well.
Finally, Theorem \ref{thm_elliptic-parametrix} shows that this implies
the existence of a parametrix for $\hat{a}$. Consequently, one can
find a self-adjoint PDO $\hat{p}=\hat{a}$ mod $\Psi^{-\infty}$ which
is invertible $C^{\infty}\left(M\right)\rightarrow C^{\infty}\left(M\right)$
and with inverse in $\Psi_{\rho}^{-m\left(x,\xi\right)}$ (see Lemma
\ref{lem_elliptic_invertible}).

\end{proof}
\normalsize

\subsubsection{$L^{2}$-continuity and quasi-compacity}

The next result, due originally to Hörmander, is very useful. It tells
us that one can take the {}``square root'' of a positive elliptic
operator. 

\vspace{0.cm}\begin{center}{\color{blue}\fbox{\color{black}\parbox{16cm}{
\begin{lem}

\label{lem_square_root}Let $p\in S_{\rho}^{m\left(x,\xi\right)}$
be an elliptic symbol satisfying $p-\mbox{Re}\left(p\right)\in S_{\rho}^{m\left(x,\xi\right)-\varepsilon}$
for some $\varepsilon>0$ and $\mbox{Re}\left(p\left(x,\xi\right)\right)>0$
for $\left\langle \xi\right\rangle \geq c$. Then, there exists $\hat{b}\in\Psi_{\rho}^{\frac{1}{2}m\left(x,\xi\right)}$
such that \[
\hat{p}-\hat{b}^{*}\hat{b}\in\Psi^{-\infty}.\]

\end{lem}
}}}\end{center}\vspace{0.cm}

\begin{proof}
\small 

Thanks to Lemma \ref{lem_elliptic_principal_symbol}, the real part
$\mbox{Re}\left(p\right)$ is elliptic and satisfies therefore $\mbox{Re}\left(p\right)\geq C\left\langle \xi\right\rangle ^{m\left(x,\xi\right)}$
for large $\left\langle \xi\right\rangle $. This implies that we
can certainly find a smooth $b_{0}\left(x,\xi\right)$ which coincides
with $\sqrt{\mbox{Re}\left(p\right)}$ for large $\left\langle \xi\right\rangle $,
i.e., outside from a compact set. Straightforward computations show
that $b_{0}\in S_{\rho}^{\frac{1}{2}m\left(x,\xi\right)}$. On the
other hand we have $\left|b_{0}\right|^{2}=\mbox{Re}\left(p\right)\mbox{ mod }S^{-\infty}$.
Then, the symbolic calculus gives \[
\hat{b}_{0}^{*}\hat{b}_{0}=\widehat{\mbox{Re}\left(p\right)}\mbox{ mod }\Psi_{\rho}^{m\left(x,\xi\right)-\left(2\rho-1\right)}=\widehat{p}\mbox{ mod }\Psi_{\rho}^{m\left(x,\xi\right)-\varepsilon},\]
where we have assumed without loss of generality that $\varepsilon\leq2\rho-1$.
The rest of the procedure is a standard iterative construction which
shows that for any $N$, there are $\hat{b}_{j}\in\Psi_{\rho}^{m\left(x,\xi\right)-j\varepsilon}$,
$j:1..N$, satisfying \[
\left(\hat{b}_{0}^{*}+...+\hat{b}_{N}^{*}\right)\left(\hat{b}_{0}+...+\hat{b}_{N}\right)=\widehat{p}\mbox{ mod }\Psi_{\rho}^{m\left(x,\xi\right)-\left(N+1\right)\varepsilon}.\]
Then, a Borel resummation (Theorem \ref{thm:borel_resummation}) yields
the result.

\end{proof}
\normalsize

The last but not the least result of this appendix is standard, since
it concerns symbols of (constant) order $0$. It is usually a way
to prove $L^{2}$-continuity of PDO's, but it yields also a way to
show that a PDO's with a {}``small'' symbol is quasi-compact, which
is the property we use in the context of Ruelle-Pollicott resonances. 

\vspace{0.cm}\begin{center}{\color{blue}\fbox{\color{black}\parbox{16cm}{
\begin{lem}

\label{lem_continuity_L2}Let $p\in S_{\rho}^{0}$ be a symbol and
denote \[
L=\limsup_{\left(x,\xi\right)\in T^{*}M}\left|p\left(x,\xi\right)\right|.\]
Then, for any $\varepsilon>0$ there is a decomposition \[
\hat{p}=\hat{p}_{\varepsilon}+\hat{K}_{\varepsilon}\]
 with $\hat{K}_{\varepsilon}\in\Psi^{-\infty}$ and $\left\Vert \hat{p}_{\varepsilon}\right\Vert \leq L+\varepsilon$.

\end{lem}
}}}\end{center}\vspace{0.cm}

\begin{proof}
\small 

The first remark is that for any $\varepsilon>0$ the operator $\left(L^{2}+\varepsilon\right)\mathbb{I}-\hat{p}^{*}\hat{p}=:\hat{q}$
is self-adjoint and in the class $S_{\rho}^{0}$, which means $q-\mathbb{R}\left(q\right)\in S_{\rho}^{-\left(2\rho-1\right)}$.
On the other hand, $q=\left(L^{2}+\varepsilon\right)-\left|p\right|^{2}$
modulo $\Psi_{\rho}^{-\left(2\rho-1\right)}$, which is positive for
large $\xi$. Therefore we can apply Lemma \ref{lem_square_root}
and obtain $\hat{b}\in\Psi_{\rho}^{0}$ such that $\hat{q}=\hat{b}^{*}\hat{b}-\hat{K}$
with $K\in\Psi^{-\infty}$. Then, for any $u\in L^{2}\left(M\right)$
one has \begin{eqnarray*}
\left\Vert \hat{p}\left(u\right)\right\Vert ^{2} & = & \left(\hat{p}^{*}\hat{p}\left(u\right),u\right)\\
 & = & \left(L^{2}+\varepsilon\right)\left\Vert u\right\Vert ^{2}-\left(\hat{b}^{*}\hat{b}\left(u\right),u\right)+\left(\hat{K}\left(u\right),u\right)\\
 & = & \left(L^{2}+\varepsilon\right)\left\Vert u\right\Vert ^{2}-\left\Vert \hat{b}\left(u\right)\right\Vert ^{2}+\left(\hat{K}\left(u\right),u\right).\end{eqnarray*}
From this follows the upper bound \begin{equation}
\left\Vert \hat{p}\left(u\right)\right\Vert ^{2}\leq\left(L^{2}+\varepsilon\right)\left\Vert u\right\Vert ^{2}+\left(\hat{K}\left(u\right),u\right).\label{eq_L2continuity_1}\end{equation}
The next step is to introduce the spectral projector $\pi_{\lambda}$
of the Laplacian $-\Delta$ on $\left(-\infty,\lambda\right]$ for
large enough $\lambda$, which will be chosen later depending on $\varepsilon$
in a suitable way. Notice that this projection is smoothing. Then,
we decompose\[
\hat{p}=\hat{p}_{\varepsilon}+\hat{r}_{\varepsilon}:=\hat{p}\left(1-\pi_{\lambda}\right)+\hat{p}\pi_{\lambda}.\]
It follows first that $\hat{r}_{\varepsilon}$ is smoothing. On the
other hand, the upper bound (\ref{eq_L2continuity_1}) yields \begin{eqnarray*}
\left\Vert \hat{p}_{\varepsilon}\left(u\right)\right\Vert ^{2} & \leq & \left(L^{2}+\varepsilon\right)\left\Vert \left(1-\pi_{\lambda}\right)u\right\Vert ^{2}+\left(\hat{K}\left(1-\pi_{\lambda}\right)\left(u\right),\left(1-\pi_{\lambda}\right)\left(u\right)\right)\\
 & \leq & \left(L^{2}+\varepsilon\right)\left\Vert u\right\Vert ^{2}+\left\Vert \hat{K}\left(1-\pi_{\lambda}\right)\right\Vert \left\Vert u\right\Vert ^{2}\end{eqnarray*}
where we have used $\left\Vert 1-\pi_{\lambda}\right\Vert \leq1$
and the Cauchy-Schwarz inequality. Finally, we show that we can make
$\left\Vert \hat{K}\left(1-\pi_{\lambda}\right)\right\Vert $ arbitrarily
small. Since $\hat{K}$ is smoothing, it is continuous $H^{s}\rightarrow L^{2}$
for any $N>0$. In particular, we can decompose \[
\hat{K}\left(1-\pi_{\lambda}\right)=\hat{K}\left(1-\Delta\right)^{N}\left(1-\Delta\right)^{-N}\left(1-\pi_{\lambda}\right)\]
 and $\left\Vert \hat{K}\left(1-\Delta\right)^{N}\right\Vert _{L^{2}}\leq C_{N}$.
On the other hand, the spectral theorem yields $\left\Vert \left(1-\Delta\right)^{-N}\left(1-\pi_{\lambda}\right)\right\Vert \leq\lambda^{-N}$.
Therefore we have showed that $\left\Vert \hat{K}\left(1-\pi_{\lambda}\right)\right\Vert \leq\frac{C_{N}}{\lambda^{N}}$
which can be made arbitrarily small, by taking $\lambda=\frac{1}{\varepsilon}$
for example. This proves $\left\Vert \hat{p}_{\varepsilon}\left(u\right)\right\Vert ^{2}\leq\left(L^{2}+2\varepsilon\right)\left\Vert u\right\Vert ^{2}$.

\end{proof}
\normalsize

\bibliographystyle{plain}
\bibliography{/home/faure/articles/articles}

\end{document}